\documentclass[preprint]{aastex}
\usepackage{longtable}
\usepackage{graphicx}
\usepackage{subfigure}
\usepackage{color}

\shorttitle{Intermediate Resolution Near-Infrared Spectroscopy of 36 late-M~Dwarfs}
\shortauthors{Deshpande et al.}

\begin{document}


\title{Intermediate Resolution Near-Infrared Spectroscopy of 36 late-M~Dwarfs}


\author{R. Deshpande\altaffilmark{1,2,3}, E. L. Mart\'in\altaffilmark{4}, M. M. Montgomery\altaffilmark{3}, M. R. Zapatero~Osorio\altaffilmark{4}, F.~Rodler\altaffilmark{5,6}, C. del Burgo\altaffilmark{7,8}, N. Phan Bao\altaffilmark{9,10},  Y. Lyubchik\altaffilmark{11}, R. Tata\altaffilmark{6},  H. Bouy\altaffilmark{4}, Y.~Pavlenko\altaffilmark{11}  }

\email{rohit@psu.edu}


\altaffiltext{1}{Center for Exoplanets and Habitable Worlds, The Pennsylvania State University, University Park, PA 16802}
\altaffiltext{2}{Department of Astronomy and Astrophysics, The Pennsylvania State University, University Park, PA 16802}
\altaffiltext{3}{University of Central Florida, Dept. of Physics, PO Box 162385, Orlando, FL, 32816-2385}
\altaffiltext{4}{Centro de Astrobiolog\'ia (CSIC-INTA), Ctra. Ajalvir km 4, 28850 Torrej\'on de Ardoz, Madrid, Spain}
\altaffiltext{5}{Institut de Ciencies de l'Espai (CSIC-IEEC), Campus UAB, Torre C5 - parell - 2a planta, 08193 Bellaterra, Spain}
\altaffiltext{6}{Instituto de Astrof\'isica de Canarias, c/ V\'ia L\'actea, s/n, E-38205 La Laguna, Tenerife, Islas Canarias, Spain}
\altaffiltext{7}{Instituto Nacional de Astrof\'isica, \'Optica y Electr\'onica (INAOE), Aptdo. Postal 51 y 216, 72000 Puebla, Pue., Mexico}
\altaffiltext{8}{UNINOVA-CA3, Campus da Caparica, Quinta da Torre, Monte de Caparica, 2825-149 Caparica, Portugal}
\altaffiltext{9}{Dept. of Physics, HCMIU, Vietnam National University Administrative Building, Block 6, Linh Trung Ward, Thu Duc District, HCM, Vietnam }
\altaffiltext{10}{Institute of Astronomy and Astrophysics, Academia Sinica, P.O. Box 23-141, Taipei 106, Taiwan}
\altaffiltext{11}{Main Astronomical Observatory of Academy of Sciences of Ukraine, Zabolotnoho, 27, Kyiv, 03680, Ukraine}

\begin{abstract}
We present observations of 36 late-M~dwarfs obtained with the KeckII/NIRSPEC in the J-band at a resolution of $\sim$ 20,000. We have measured projected rotational velocities, absolute radial velocities, and pseudo-equivalent widths of atomic lines. 12 of our targets did not have previous measurements in the literature.

For the other 24 targets, we confirm previously reported measurements. We find that 13 stars from our sample have $v$~sin~$i$ below our measurement threshold (12 km s$^{-1}$) whereas four of our targets are fast rotators ($v$~sin~$i$ $>$ 30 km s$^{-1}$). As fast rotation causes spectral features to be washed out, stars with low projected rotational velocities are sought for radial velocity surveys.

At our intermediate spectral resolution we have confirmed theidentification of neutral atomic lines reported in \citet{mclean07}. We also calculated pseudo-equivalent widths (p-EW) of 12 atomic lines. Our results confirm that the p-EW of K I lines are strongly dependent on spectral types. We observe that the p-EW of Fe I and Mn I lines remain fairly constant with later spectral type. We suggest that those lines are particularly suitable for deriving metallicities for late-M dwarfs.
\end{abstract}


\section{Introduction} 

Radial velocity studies of early bright M dwarfs have yielded planets, including a planetary system \citep{marcy99} and rocky planets \citep{rivera05, udry07, mayor09}. Though M dwarfs present themselves as promising candidates for rocky planet searches in the habitable zone, the effort required to measure precise radial velocities are often thwarted by their higher projected rotational velocities ($v$~sin~$i$ $>$ 30 km s$^{-1}$) and stellar activities.

Radial velocity measurement precision is limited by stellar rotation. An increase in stellar rotation causes narrow deep lines to become broad, shallow, and blended. Such lines reduce available radial velocity information, thereby reduce the precision. \citet{reiners10b} find that approximately 50\% of their sample of 63 M dwarfs with spectral types M7-M9.5 show projected rotational velocities greater than 10 km s$^{-1}$. A comprehensive study of projected rotational velocities of early- to mid-M dwarfs suggests an increasing trend in projected rotational velocity with later spectral type \citep{jenkins09}.

The relation between stellar activity and projected rotational velocity is well established in literature \citep{noyes84, delfosse98a, pizzolato03}. This relation also holds true for stars at the end of the main sequence \citep{mohanty03}. Stellar activity produces stellar spots. These spots distort the line profiles of the stellar absorption lines that are critical for radial velocity measurement. This distortion leads to a change in the bisectors of the absorption lines. Such lines can impersonate an unseen companion thus resulting in a false detection \citep{henry02}. An extensive simulation work on stellar spots by \cite{reiners10a} suggests that spots can cause radial velocity shifts of 100 m s$^{-1}$. 

A comprehensive study of stellar activity of M dwarfs (West et al. 2004) indicates that the fraction of active stars increases from early-M dwarfs (10$\%$) to late-M dwarfs (75$\%$). Hence, late-M dwarfs are likely to be fast rotating and active. With the upcoming infrared radial velocity surveys dedicated to search for planets around M dwarfs, such as HPF \citep{mahadevan10} and CARMENES \citep{quirrenbach10}, precise measurements of stellar rotation among late-M~dwarfs becomes crucial.

Measurement of radial and projected rotational velocities requires a good understanding of lines in the stellar atmosphere. Over the last seven years, identification and characterization of neutral atomic lines have been done at low (R $\sim$ 2000) \citep{cushing05} and intermediate resolutions (R $\sim$ 20,000) \citep{mclean07}. However, both these studies used a small sample of M dwarfs and calculated pseudo-equivalent width (p-EW) of a few atomic lines. With 36  late M~dwarfs (M5.0 - M9.5), we have more than doubled the number of objects for which p-EW of 12 neutral atomic lines have been measured.

In this paper, we verify the identification of neutral atomic lines using the Vienna Atomic Line Database \citep{kupka02}. We calculate rotational and absolute radial velocities (in a companion paper, we reported relative radial velocities for stars with multiple observations \citet{rodler12}), and measure p-EWs of 12 neutral atomic lines.

The paper is organized as follows: In $\S$2, we describe our sample, compare it with existing data,  provide our instrumental setup for observations, and data reduction procedure. In $\S$3, we list the results of our observations that we analyze and discuss in $\S$4. In $\S$5, we summarize our work and provide conclusions.

\section{The Sample, Observations and Data Reduction}

\subsection{Sample}
Our sample of targets is listed in Table 1. Furthermore the table also provides 2MASS and alternate target names, spectral types (determined in the optical), exposure times, observation dates, J-band magnitudes, and reference to spectral types. The J-band magnitude of the stars vary from $\sim$ 7 to 13 mag implying spectrophotometric distances of up to 40 pc \citep{phan08}. Figure~1 shows the distribution of our targets in terms of spectral types. 

Most of our stars were not known to be binaries with the exception of LP349-25 and 2MASS J2206-2047. LP349-25 was first recognized as a nearby M dwarf  \citep{gizis02} but was later revealed as a close binary system with an angular separation of 0.125 $\pm$ 0.01" \citep{forveille05}. 2MASS J2206-2047 was also found to be a binary system with an angular separation of 0.168 $\pm$ 0.007" \citep{close02}.

\subsection{Observations}

Our sample of 36 M dwarfs were observed with the near-infrared echelle spectrograph NIRSPEC \citep{mclean98} installed at the Keck II 10-m telescope. NIRSPEC is a cryogenic cross-dispersed echelle spectrograph with an ALADDIN-3 InSb 1024$\times$1024, 27 $\mu$m detector. Our observing program was granted one and half nights for semester 2007A and two nights for semester 2007B, giving us a total of 3.5 nights of NASA Keck time (see Table~1 with log of observations). Observing conditions were good except for the nights of April 30 and October 26, which were affected by computer crashes, and for the nights of December 22-23, which were due to bad weather.

We used the echelle mode with the NIRSPEC-3 (J-band) filter. The echelle angle was fixed at $\sim$ 63$^{\circ}$ with the cross-disperser at $\sim$34$^{\circ}$. The entrance slit was set at a width of 0.432" and a length of 12''. This setup provided a wavelength range of ($\sim$ 1.15 - 1.36 $\mu$m) split into 11 echelle orders (\textit{m} = 66 - 56). For echelle numbering scheme we refer to \cite{mclean07}. Echelle orders 66, 57, and 56 are heavily contaminated by telluric lines while no atomic lines are observed in orders 62 and 63. Therefore, for this paper we made use of six echelle orders (\textit{m} = 65, 64, 61 - 58) which cover an effective wavelength ranges of  (1.1649 - 1.20011 $\mu$m) and (1.24081 - 1.32370 $\mu$m). The nominal dispersion ranged from 0.167 \AA $\ $ at blue wavelengths, to 0.191 \AA $\ $at red wavelengths, and the final resolution elements ranged from (0.55 - 0.70 \AA). The resolving power of our spectra ranged from (17,800 - 22,700) \AA $\ $ \citep{zapatero06, lyubchik07}. 

The individual exposure times, based on the J-band magnitude of the targets, ranged from 20 to 300 seconds per co-add. The targets and the telluric standards (spectral type A stars) were nodded in the A-B format, where A and B are the target positions along the slit that are substantially separated ($\sim$ 7''), in order to subtract sky background. Flats were taken frequently throughout the observing run. We took spectra of a ThAr lamp immediately after every target to ensure accurate wavelength calibration. In order to remove atmospheric telluric lines, featureless stars in the near-IR (spectral types A0-A2) were observed close to the target both in time and in position of the sky to ensure similar airmass.

\subsection{Data Reduction}

The data reduction procedure in this paper is similar to that provided in \citep{zapatero06}. The bias and dark current subtraction, flat-fielding, wavelength calibration, and telluric lines removal are carried out using the \tt{echelle} \rm{} package within the Image Reduction and Analysis Facility (IRAF)$^{\footnote{IRAF is distributed by the National Optical Astronomy Observatories, which are operated by the Association of Universities for Research in Astronomy, Inc., under cooperative agreement with the National Science Foundation}}$ \citep{tody93, tody86}. The flats taken at the same instrumental setup are used to flat-field the images. The sky subtraction on the spectra is performed by subtracting the two nodded images. The spectra is then wavelength calibrated using the emission lines of ThAr lamp. The lines are identified prior to wavelength calibration using the National Institute of Standards and Technology (NIST)$^{\footnote{\url{http://physics.nist.gov/PhysRefData/ASD/lines_form. html}}}$ line database. The fits to these lines are made using a third order Legendre polynomial along the dispersion axis and second-order perpendicular to it. The mean rms for the fit is $\sim$ 0.016 \AA $\ $or 0.38 km/s when calculated from the central wavelength of 12485 \AA$\ $. The wavelength calibrated spectra is then normalized using the \tt{continuum}\rm{} package in IRAF. The reduction of telluric standards is performed in an identical manner. Furthermore, a strong hydrogen line at 1.282 $\mu$m and few weaker intrinsic stellar lines present in the telluric standard are removed manually by fitting the line profile using the task \tt{splot}\rm{} in IRAF. The wavelength calibrated and normalized target spectra are then divided by the normalized wavelength calibrated telluric standard. The output spectra, A-B, is then free of telluric lines. The whole process is repeated on the other subtracted nodded image, B-A.  In order to increase the signal-to-noise ratio (SNR), the spectra of nodded target (i.e A-B and B-A) are combined using the \tt{scombine} \rm{}package within IRAF. 

Since FeH and H$_{2}$O molecular lines are present all over our spectral range, we present an approximate SNR values based on photon noise calculations. Using the instrument gain factor and normalized photon counts, as exposure times of the targets varied, we compute the SNR for all of our targets. These values are listed in column sixth column of Table 2.

Figure 2 shows a sample of 10 targets that were reduced using the method described above. This figure illustrates the spectra of these targets as seen in six orders. For clarity the stellar spectra are stacked and separated by a constant value. The vertical axis on the right of each plot notes the spectral type. The strong and weak neutral atomic lines are indicated by dotted lines in the figure. 

\section{Results}

\subsection{Projected Rotational Velocity}

The projected rotational velocities ($v$~sin~$i$) presented in this paper are measured by cross-correlating the spectra of our targets with the spectra of slowly rotating template of similar spectral type and observed with the same instrumental configuration. Procedures used to calculate projected rotational velocities are described in the literature  \citep{tinney98, mohanty03, white03, bailer04}. With the assumption that the line profile is primarily dominated by rotation, projected rotational velocities are calculated by measuring the width of the function obtained from the cross correlation of the target's spectrum with a slowly rotating template.

To calibrate this relationship, we artificially broaden the template spectrum for a range of velocities (10-50~km~s$^{-1}$, in steps of 5 km s$^{-1}$ and limb-darkening parameter, $\epsilon$ = 0.6 \citep{claret98}. We employ the line rotation profile given by \citet{gray05}. A spectral order in a velocity broadened template is cross correlated with the same order in the original template. A Gaussian is fit to the cross correlation function (CCF) and its width is measured. This width is associated to the velocity by which the original template was broadened for a given order. The process is repeated for the range of velocities and orders. A correlation table between the widths and the velocities is thus created. As a final step, the CCF width measured by cross-correlating a target and the original template is associated with a projected rotational velocity from the table. The process is repeated for all orders. The mean of the associated velocities is taken to be the projected rotational velocity of the star. The errors in the measurement are calculated from the standard deviation of the mean. In our study we employed six orders. Table 2 lists the $v$~sin~$i$ of our targets. Furthermore, Table 2 also lists the targets by their 2MASS and alternate names, spectral types, literature $v$~sin~$i$, and their references.

Ideally, non-rotating stars should be used. However, all stars rotate and the utilization of synthetic spectra is outside the scope of this paper. Hence, we have used templates that have published $v$~sin~$i$ values well below the resolution of our spectra, which is estimated to be $\sim$ 12 km s$^{-1}$. For all our targets for which we find $v$~sin~$i$ at or below 12 km s$^{-1}$, we report an upper limit.

As our sample consists of targets with a range of spectral types, there is a possibility of spectral mismatch when computing projected rotational velocities with a single template. To see whether spectral type and $v$~sin~$i$ of the template affect the calculations of projected rotational velocities, we considered two slowly rotating stars (GJ406 $\&$ vB10). The reported projected rotational velocities of these two stars are: $v$~sin~$i$ $< $ 2.5 km s$^{-1}$ \citep{delfosse98a} and $v$~sin~$i$ = 6.5 km s$^{-1}$ \citep{mohanty03}, respectively. We measured $v$~sin~$i$ of all stars from our sample using these two templates. Figure 3 illustrates this effort. The measured $v$~sin~$i$ are within each other errors. However, we obtain higher measurement precision when the target's spectral type is closer to that of a template. Hence we have employed GJ406 (M6.0) as a template to compute $v$~sin~$i$ for stars with spectral types M5.0 - M6.5 and vB10 (M8.0) for M7.0 - M9.5. 

\subsection{Heliocentric Radial Velocity}

Absolute radial velocities are determined by cross correlating the target spectra with the radial velocity template spectrum: The telluric-free normalized spectra of each object and vB10 are cross correlated using the \tt{fxcor}\rm{} package in IRAF. As most target spectra have high SNR, the peak for each CCF is narrow and well defined. A gaussian is fit to the CCF peak and the radial velocity is determined.

As our sample spans a wide range of spectral types, we tested the same two templates, GJ406 and vB10, as we did in the previous section. GJ406 is known to have a low projected rotational and absolute radial velocity. However, vB10 has been extensively used in the literature as a radial velocity template with a reported velocity of $\sim$~35~km~s$^{-1}$,  \citep{tinney98, martin99, basri06, zapatero07}. Unlike GJ406, vB10 is known to flare occasionally \citep{linsky95}. However, radial velocity measurements of vB10 over the last 12 years have shown no significant variation. vB10 had been suspected of harboring a gas-giant planet \citep{pravdo09, zapatero09} but high precision radial velocity monitoring \citep{bean10, ang10, rodler11} have refuted such claim. Thus we consider GJ406 (19.0~km~s$^{-1}$; \citet{mohanty03}) and vB10 (34.37~$\pm$~0.3~km~s$^{-1}$; \citet{zapatero09}) as radial velocity templates.

The systematic errors or zero-point shifts that are likely to be present in the wavelength calibration of our data affect the measurement of absolute radial velocities. The observed velocity shifts in the telluric lines are $\sim \pm$ 10 m s$^{-1}$  \citep{seifahrt08} which are below 10 times the velocity precision we can achieve with NIRSPEC \citep{zapatero09}. Therefore, at these precisions, telluric lines appear to be stationary. These lines, present in the target spectrum, become excellent calibration tool for measuring instrumental zero-point shifts. We cross-correlate the heavily contaminated telluric orders in the target spectrum ( e.g. orders 59 $\&$ 58) with a synthetic telluric spectrum \citep{rothman09} at the same resolution. The velocity shifts between the two spectra give the zero-point. The value is then subtracted from the radial velocity of the target. The radial velocities are corrected for barycentric motion. This procedure is repeated for each of the 6 orders. After applying the zero-point correction to the measured velocity in each order, the mean of radial velocities values is taken as the heliocentric radial velocity for the target while the standard deviation of the mean is estimated as the error in determination of radial velocities. We find that (Figure 3, left panel) both templates, GJ406 and vB10, yield measurements that are within each other errors. Figure 4 (left panel) compares absolute radial velocities of targets from our sample with those in the literature. We find that the radial velocity measurement precision is better for a target of spectral type that is closer to the template. Therefore, we adopt GJ406 (M6.0) as a template for stars with spectral types M5.0 - M6.5 and vB10 (M8.0) for M7.0 - M9.5. Table 3 lists absolute radial velocity for our sample. In addition, for completeness, the table also lists trigonometric parallaxes and spectrophotometric distances to the stars. Literature radial velocities are also listed.

\subsection{Atomic Line Identification and Pseudo-Equivalent Width Measurements}

We employed the Vienna Atomic Line Database \citep{kupka02} to try to identify new lines in the spectra. We set the following stringent criteria to search for new lines: (1) Lines with oscillator strengths (log(gf) $<$ -2.0) were excluded to ensure that the most probable transitions are chosen; (2) if a line from the spectra is within 0.6~\AA$ \ $(minimum separation that can be resolved) of a line in the database, then that wavelength value is selected; and (3) we compare the positions of all lines in the target spectra with those positions of lines in the synthetic spectra that have been generated using the WITA6 program \citep{pavlenko00} and the NextGen model atmosphere structures \citep{hau99} for cool stars \citep{lyubchik04}.  From this procedure, we conclude that new lines other than those identified by \citet{mclean07} were identified.

If temperatures drop below 4000 K, molecules form in the atmosphere of M~dwarfs \citep{burrows03}. Unlike atoms, molecules have a large number of energy levels due to their rotational and vibrational states. Such energy levels lead to band structures where numerous individual transitions take place \citep{ten07}. These individual transition features dominate the M dwarf's spectra at infrared wavelengths. This severe blanketing of lines hinders identification of the continuum. Therefore, the traditional method of calculating equivalent widths (EW) cannot be utilized.  Instead, we search for the pseudo-continuum which is formed by molecular absorptions \citep{martin96, pavlenko95}. EW measurements at these pseudo-continuum are termed p-EWs \citep{zapatero02}. 

For the determination of the flux envelope or pseudo-continuum, we use the relation $\sigma$ = 100$\times$SNR$^{-1}$. Such envelope is dependent on the SNR of the spectra. We term such continuum as pseudo-continuum. Once the pseudo-continuum is determined, the pseudo-equivalent width is measured by direct integration of the observed line in the spectra. We calculate the uncertainty in p-EW by using the relation given by \citep{cayrel88} which estimates the uncertainty of p-EW as a function of spectral quality \citep{stetson08}.  As the true continuum cannot be determined precisely, we obtain a lower limit to the estimated uncertainty in p-EW measurements.

\section{Analysis and Discussion}

\subsection{Projected Rotational and Absolute Radial Velocities}

We have identified 13 targets in our sample with $v$~sin~$i$ below our detection threshold ($\sim$ 12 km s$^{-1}$), while four stars (LP44-1627, LP349-25, 2MJ0004-1709, and 2MASSJ1835+3259) show velocities greater than 30 km s$^{-1}$. Figure 4 (right panel) compares $v$~sin~$i$ values of our sample with those in literature (\citet{stauffer86, marcy92, delfosse98b, gizis02, mohanty03, bailer04, fuhrmeister04, jones05, reiners07, west08, reiners10b, jenkins09}). Our results indicate low residuals for stars with projected rotational velocities below 30 km~ s$^{-1}$. Though, the three fast rotating stars show larger velocity dispersion their measurements are within 3-$\sigma$ of the literature values.

The measurement of rotational velocity of the two components of LP349-25 using NIRPSEC and LGS AO \citep{konopacky12} in conjunction suggests that LP349-25A (M8.0) has a velocity of 55 $\pm$ 2 km s$^{-1}$ while the fainter LP349-25B has a velocity of 83 $\pm$ 3 km s$^{-1}$. Our measurement for the combined spectra is closer to the primary than the secondary. As the secondary is faint, the projected rotational velocity is dominated by the primary. This is particularly evident in works of \citet{reiners10a} where, using R $\sim$ 31,000, their measured projected rotational velocity is also closer to that of the primary. The same scenario is observed for the case of 2M2206-2047, where the projected rotational velocities of the two components, A and B are 19 $\pm$ 2.0 km~ s$^{-1}$ and 21 $\pm$ 2.0 km~ s$^{-1}$ , respectively. We find the velocity of the combined spectrum to be 22.2 $\pm$ 2.0 km~ s$^{-1}$. However, as both components are M8.0, their contribution is similar. 

\subsection{Neutral Atomic Lines in M dwarfs}

Figure 2 plots a sample of our reduced spectra by NIRSPEC echelle orders with order 65 in the upper left panel and order 58 in the lower right panel. Atomic absorption lines are identified by dashed lines. Spectral types are listed on the right side of each plot. These spectra are similar to those shown in \citet{mclean07}, however with larger sample size each panel shows a continuous transition from M5.0 to M9.5 in steps of half spectral type. Tables 4 - 6 list the p-EW of 12 neutral lines.

The K~I doublet at 11690 \AA$ \ $and the K~I triplet 11771 \AA$ \ $in order 65 (Figure 2, top left panel) show increasing line width with later spectral type. By M9.5 the width of the K~I triplet increases to the extent that it blends into a single line. The widening of lines is induced by collisional broadening with H$_2$ molecules in the atmospheres of M~dwarfs \citep{mclean07}. We also find that the K~I lines in order 61 have a similar trend like those in order 65. These results shown in Figure 5 agree with the trend found by \citet{mclean07}, and \citet{cushing05}. A weak Fe~I line in order 65 is observed at longer wavelengths. This line weakens with later spectral type and by M9.5 it is almost indiscernible.

The modeled spectra of late M~dwarfs by \citet{lyubchik04} and \citet{rice10}, indicate a presence of weaker Fe~I and Ti~I lines in order 65. We identified their positions, but found them to be heavily blended and hence are not clearly distinguishable in our spectra. Better spectral resolution is required in order to distinguish these lines.

The Fe~I lines (doublet 11886 \AA$ \ $, 11887 \AA$ \ $ and a singlet 11976  \AA$ \ $) dominate order 64 (Figure 2, top right panel). At M5.0, Fe~I lines are seen as narrow sharp absorption lines. However, with later spectral type, the Fe~I line at shorter wavelength broadens and blends with surrounding lines. Our results agree with those in \citet{cushing05} within the errors. A quantitative comparison is discussed in the next section.

Order 64 also contains the Mg I line at 11828 \AA$ \ $and Ti~I line at 11893 \AA. Both lines are weak and weaken with later spectral type. The Ti~I line disappears around M9 while the Mg I line becomes indistinguishable around M8.5. The Mn~I line (12899 \AA) in order 59 (Figure 2, bottom left panel) maintains its strength with later spectral type while the two Ti~I lines are observed to weaken with later spectral types. As seen in the figure, the Mn~I line maintains its sharp and narrow feature with later spectral type.

The two Al~I lines (13127 \AA$ \ $, 13150 \AA) in order 58 (Figure 2, bottom right panel) weaken with later spectral types. The Na I line at 12682 \AA$ \ $ (Figure 2, middle right panel) also shows a decline in strength with later spectral type. By M8.5 the Na I is completely indistinguishable. 

\subsection{Pseudo-Equivalent Widths}

As a trend of increasing width with later spectral type is qualitatively observed in the neutral atomic lines in Figure 2, a quantitative comparison is desired. We calculate the p-EWs of neutral atomic lines shown in Figure 2.  Figures 5-7 plot the computed p-EWs for our targets with respect to the spectral types.

Comparing our data (open circles) with those from \citet{cushing05} in figures 5 and 6, we see that Cushing and colleagues' values (filled diamonds) are within our expected errors. For clarity, in figures 5 and 6, \citet{cushing05} values have been shifted to the right by 0.1 spectral types. However, p-EW values by \citet{mclean07} for two late M dwarfs do not match with ours or with those of \citet{cushing05}. These values are not plotted as they are well beyond the range of the plot.

In each panel of Figures~5-7, a linear regression fit to the data is shown as a short-dash line. The linear relation including its y-offset and slope are listed in Table 7 for the 12 neutral atomic lines. These relations quantitatively illustrate the variation of each atomic line with respect to spectral type. The top two panels of Figure 5 show an increasing trend of p-EWs with later spectral type for K~I lines. Figure~5 compares p-EW of our sample  with those from \citep{cushing05}. We find that our p-EW measurements agree with theirs.

The p-EW calculations of Fe~I lines in figure 6 top panel suggests small variation with spectral type. However, the Al~I line shows a decreasing trend. Mn~I line along with two Ti~I lines, are observed in Order 59. As seen in Figure 2, the Mn~I line maintains its narrow sharp feature with later spectral type. This line has been studied in greater detail by \citet{lyubchik07}. They model the Mn~I line in five ultracool dwarfs ranging from M6 to L0. Using the WITA6 program \citep{pavlenko00} for the NextGen model structure they have been able to fit the Mn~I line, and their p-EW calculations carried out using the DECH20 package \citep{galazutdinov92} on three dwarfs between M6 and M9 indicate a variation of 2 m\AA. Our results confirm that the Mn~I line is not sensitive to variation in spectral type.

\section{Final Remarks}

We have observed 36 M~dwarfs with the NIRSPEC instrument in the J-band at a resolution of $\sim$ 20,000. We have measured projected rotational velocities of these stars. Never previously reported absolute radial and projected rotational velocities of 12 targets are provided. For other targets, we confirm previously reported velocity measurements. We find that 13 stars from our sample have $v~\sin~i$ below our measurement threshold (12 km s$^{-1}$) whereas four of our targets are fast rotators ($v~\sin~i$ $>$ 30~km~s$^{-1}$). As fast rotation causes spectral features to be washed out, stars with low projected rotational velocities are sought for radial velocity surveys.

At our intermediate spectral resolution we have confirmed and the identification of neutral atomic lines reported in \citet{mclean07}. We also calculated pseudo-equivalent widths (p-EW) of 12 strong and weak neutral atomic lines. Our results confirm that the p-EW of K I lines are strongly dependent on spectral types. We observe that the p-EW of Fe I and Mn I lines remain fairly constant with later spectral type. 

The absolute radial velocities and projected rotational velocities of our targets were calculated using GJ406 and vB10 as reference. Our results agree well with those in the literature for the stars in common. 

\acknowledgements
We would like to thank the anonymous referee for valuable comments and suggestions. We also like to thank the observing staff at the W.M. Keck Observatory with the observations and acquisition of data throughout 2007. We thank the Hawaiian ancestry who allowed us to observe from their sacred mountain.  We also acknowledge the authors who have contributed various atomic linelists to {\scriptsize VALD} database. R.D would like to thank Mike Fitzpatrick and Francisco Valdes for answering queries in {\scriptsize IRAF}. The research in this paper made use of {\scriptsize SIMBAD} database operated at {\scriptsize CDS}, Strasbourg, France. The support for this project comes from the award (No. 1326479) issued by Jet Propulsion Laboratory (JPL)/ National Aeronautics and Space Administration (NASA) and Florida Space Initiative (SRI). R.D would also like to acknowledge the Center for Exoplanets and Habitable Worlds for supporting his research. The Center for Exoplanets and Habitable Worlds is supported by the Pennsylvania State University, the Eberly College of Science, and the Pennsylvania Space Grant Consortium. E.L.M was supported by the CONSOLIDER-INGENIO GTC project and by AYA2011-30147-C03-03 project of the Spanish Ministry of Science.Work of Y.P and Y.L. was partially supported by the ``Cosmomicrophysics'' program of the Academy of Sciences of Ukraine and EU PF7 Marie Curie Initial Training Networks (ITN) ROPACS project (GA N 213646). Work of C. del B. was partly funded by the Funda\c{c}\~ao para a Ci\^encia e a Tecnologia (FCT)-Portugal through the project PEst-OE/EEI/UI0066/2011

{\it Facilities:} \facility{Keck II}, \facility{NIRSPEC}

\clearpage




\clearpage

\clearpage

\begin{deluxetable}{c c c c c c c }
\tablewidth{0pt}
\tablecaption{Keck/NIRSPEC Observing Log}
\tablehead{
\colhead{Object}           & \colhead{Alt.Name}      &
\colhead{Sp. Type}          & \colhead{$T_{exp}$ (sec.)}  &
\colhead{Obs. Date}          & \colhead{J (mag.)} & \colhead{SpT. Ref.}
}
\tabletypesize
\scriptsize{}
\startdata
2MASS J00045753-1709369 	&-	& M5.5		&	100		&	2007 Jun. 24    &  10.99		&      (1)    	\\						
	&                   		& M5.5		&      100		&      2007 Jun. 25    &  		& 					\\ 			
				&                   		& M5.5 		&      300	        &      2007 Oct. 26    & 		& 					\\ 	
2MASS J00064325-0732147 	&     GJ1002      	& M5.5		&	70		&	2007 Jun. 25   	&  8.32 	& 	(2)	\\					
2MASS J00130931-0025521		&	-			& M6.0		&	200		&	2007 Jun. 25	&  12.17	&	(3)	\\	
2MASS J00275592+2219328	&    LP349-25  		& M8.0		&	120		&	2007 Jun. 24   	&  10.61	& 	(4)	\\				
                                      &                    	&M8.0  		&      120	        &      2007 Jun. 25   	&		& 					\\ 					
                                      &                    	&M8.0  		&      300	        &      2007 Oct. 26    	&		& 					\\ 					
2MASS J01095117-0343264	&  LP647-13		&M9.0		&	300		&	2007 Oct. 26 	&  11.69	&	(5)	 \\				
				&				&M9.0		&	300		&	2007 Oct. 27	&		&   \\
2MASS J01400263+2701505	&	-      &M8.5		&	300		&	2007 Oct. 26	&  12.49	& 	(6)	\\		
2MASS J02141251-0357434		&	LHS1363		&M6.5		&	300		&	2007 Oct. 27    	&  10.48	 & 	(7)	\\		
				&				&M6.5		&	300		&	2007 Oct. 27 	&		&  \\
2MASS J02532028+2713332	&	-			&M8.0		&	300		&	2007	 Oct. 26	&  12.50	& 	(6)	\\	
2MASS J03205965+1854233	&	LP412-31		&M8.0		&	300		&	2007 Oct. 26	&  11.76	& 	(8) \\				
				&				&M8.0		&	300		&	2007 Oct. 27	&		& \\
2MASS J03505737+1818069	&	LP413-53		&M9.0		&	300		&	2007 Oct. 26	&  12.97	& 	(6)	\\			
				&				&M9.0		&	300		&	2007 Oct. 27	&		& \\
2MASS J03510004-0052452		&	LHS1604		&M8.0		&	300		&	2007 Oct. 26	&  11.30	& 	(4)	\\		
2MASS J03542008-1437388		&	-			&M6.5		&	300		&	2007 Oct. 27	&  11.34	& 	(9)	\\	
2MASS J04173745-0800007  		&	-			&M7.5		&	300		&	2007 Oct. 26	&  12.18	&	(10) \\		
2MASS J04351612-1606574		&	LP775-31 	&M7.0		&	300		&	2007 Oct. 27	&  10.41	& 	(4)	\\			
2MASS J04402325-0530082	&	LP655-48		&M7.0		&	300		&	2007 Oct. 26	&  10.66	& 	(4)	\\			
2MASS J04433761+0002051	&	-			&M9.0		&	300		&	2007 Oct. 26	&  12.51	& 	(10) \\			
2MASS J05173766-3349027	&	-				&M8.0		&	250		&	2007 Oct. 27 	&  12.00	& 	(10) \\		
2MASS J07410681+1738459 	&	LHS1937		&M7.0		&	300		&	2007 Oct. 27	&  12.01	& 	(11)	\\			
2MASS J10562886+0700527	&	GJ406		&M6.0		&	120		&	2007  Apr. 30	&  7.09	& 	(12)	\\			
				&				&M6.0		&	30		&	2007 Dec. 22	&		& \\
				&				&M6.0		&	30		&	2007  Dec. 23  	&		& \\
2MASS J12185939+1107338	&	GJ1156		&M5.0		&	120		&	2007 Apr. 30	&  8.52	&	(12) \\				
				&				&M5.0		&	300		&	2007 Dec. 22   	&		& \\
2MASS J12531240+4034038	&	LHS2645		&M7.5		&	300		&	2007 Apr. 30	& 12.19	&	(13) \\				
2MASS J14563831-2809473		&	LHS3003		&M7.0		&	200		&	2007 Apr. 30	&   9.97	& 	(8) \\			
2MASS J15460540+3749458	&	-			&M7.5		&	300		&	2007 Apr. 30	& 12.44	&	(4)	\\		
				&	-			&M7.5		&	300		&	2007 Jun. 24	&		& \\
2MASS J15524460-2623134	&	LP860-41		&M6.0		&	300		&	2007 Apr. 30	&  10.26	& 	(1) \\				
2MASS J16142520-0251009 	&	LP624-54		&M5.0		&	200		&	2007 Apr. 30 	&  11.30	& 	(14) \\				
				&				&M5.0		&	200		&	2007 Jun. 25	&		& \\
2MASS J1733189+463359 		&	-			&M9.5		&	300		&	2007 Jun. 24	&  13.24	& \\			
2MASS J17571539+7042011	&	LP44-162		&M7.5		&	120		&	2007 Jun. 24	&  11.45	&	(13) \\				
				&				&M7.5		&	250		&	2007 Jun. 25 	&		& \\
2MASS J18353790+3259545	&	-			&M8.5		&	120		&	2007 Jun. 24 	& 10.27	& 		(15) \\		
				&				&M8.5		&	120		&	2007 Jun. 25 	&		& \\
2MASS J18432213+4040209	&	LHS3406		&M8.0		&	300		&	2007 Jun. 24 	& 11.30		& 	(4) \\			
				&				&M8.0		&	300		&	2007 Jun. 25 	&		& \\
2MASS J19165762+0509021	&	vB10		&M8.0		&	120		&	2007 Jun. 25 	& 9.91	& 	(13) \\				
2MASS J22062280-2047058		&	-			&M8.0		&	300		&	2007 Jun. 24	& 12.37	&	(16)  \\		
				&	-			&M8.0		&	300		&	2007 Jun. 25	& 		& \\
2MASS J22081254+1036420 	&	RXJ2208.2	&M5.0		&	120		&	2007 Jun. 24	&10.60	& 	(17) \\					
				&				&M5.0		&	120		&	2007 Jun. 25	&		& \\
2MASS J23312174-2749500		&	-			&M7.0		&	200		&	2007 Jun. 24	& 11.65	&	(18)  \\		
				&	-			&M7.0		&	200		&	2007 Jun. 25	&		& \\
2MASS J23334057-2133526		&	LHS3970		&M5.0		&	200		&	2007 Jun. 24	& 10.26	& 	(1) \\			
				&				&M5.0		&	200		&	2007 Jun. 25	&		& \\
2MASS J23415498+4410407	&	GJ905		&M6.0		&	20		&	2007 Jun. 25	& 6.88	& 	(12) \\				
	&				&M6.0		&	120		&	2007 Oct. 27	&		&	(13) \\		
2MASS J23494899+1224386	&	LP523-55		&M8.0		&	300		&	2007 Oct. 27	& 12.60	&	(4)  \\

\enddata
\tablerefs{
(1) Crifo et al. 2005; (2) Leggett 1992; (3) West et al. 2008; (4) Gizis et al. 2000; (5) Cruz \& Reid 2002; (6) Allen et al. 2007
(7) Reid et al. 2002; (8) Kirkpatrick et al. 1995; (9) Caballero 2007; (10) Schmidt et al. 2007; 
(11) Gizis  \& Reid 1997; (12) Basri 2001; (13) Kirkpatrick, D. et al. 1991; (14) Phan Bao, N. et al., 2006; (15) Reid, I.  et al. 2003;
(16) Cruz et al.  2003; (17) Zickgraf et al. 2005; (18) Reiners, A., \& Basri, G. 2009 }
\end{deluxetable}

\clearpage

\begin{deluxetable}{c c c c c c c c}
\tablewidth{0pt}
\tablecaption{Rotational Velocity and SNR}
\tablehead{
\colhead{Object}           & \colhead{Alt.Name}      &
\colhead{Sp. Type}          & \colhead{v sin $i$ (km sec$^{-1}$)}  &
 \colhead{Lit. v sin $i$ (km sec$^{-1}$)} & 
\colhead{SNR}    & \colhead{Ref} 
} 
\addtolength{\tabcolsep}{-6pt}
\tabletypesize
\scriptsize{}
\startdata
2MASS J12185939+1107338	&	GJ1156		&	M5.0	&	17.4 $\pm$ 3.0		&	17			    		&	237 	& (a)	\\ 	
2MASS J16142520-0251009	&	LP624-54		&	M5.0	&	$<$12			&	-			            	&	58 	&	\\ 	
2MASS J22081254+1036420	&	RXJ2208.2	&	M5.0	&	18.6 $\pm$ 2.0		&	-			            	&	120	&	\\ 	
2MASS J23334057-2133526	&	LHS3970		&	M5.0	&	$<$12			&	-			            	&	90	&	\\ 	
2MASS J00045753-1709369	&		-		&	M5.5	&	30.8 $\pm$ 2.0		&	-			            	&	76	&	\\ 	
2MASS J00064325-0732147	&	GJ1002		&	M5.5	&	$<$12			&	$\le$ 3		            	&	273	& (a)	\\ 	
2MASS J00130931-0025521	&		-		&	M6.0	&	17.4 $\pm$ 3.0		&	-			            	&	43	&	\\ 	
2MASS J15524460-2623134	&	LP860-41		&	M6.0	&	16.1 $\pm$ 3.0		&	-			            	&	83	&	\\ 	
2MASS J23415498+4410407	&	GJ905		&	M6.0	&	 $<$ 12			&	$\le$ 1.2	            		&	80	& (b)\\ 	
2MASS J02141251-0357434	&	LHS1363		&	M6.5	&	$<$12			&	-			            	&	105	&\\ 	
2MASS J03542008-1437388	&		-		&	M6.5	&	23.2 $\pm$ 2.0		&	-			            	&	69	&\\ 	
2MASS J10562886+0700527	&	GJ406		&	M6.5	&	template			&	$\le$ 3		            	&	399	& (b)	\\ 
2MASS J04351612-1606574	&	LP775-31		&	M7.0	&	12.1 $\pm$ 3.0		&	-			            	&	104	& 	\\ 
2MASS J04402325-0530082	&	LP655-48		&	M7.0	&	19.6 $\pm$ 2.0		&	16.2 $\pm$ 2.0          	&	103	& (c)	\\ 
2MASS J07410681+1738459	&	LHS1937		&	M7.0	&	13.0 $\pm$ 3.0		&	10.0$\pm$ 2.0           	&	52	& (c)	\\ 
2MASS J14563831-2809473	&	LHS3003		&	M7.0	&	$<$12			&	5.0 $\pm$ 2.0           	&	108	& (c)	\\ 
2MASS J23312174-2749500	&		-		&	M7.0	&	$<$12			&	9.0 $\pm$ 2.0           	&	62	& (c)	\\ 
2MASS J04173745-0800007	&				&	M7.5	&	$<$12			&	7.0 $\pm$ 2.0           	&	45	& (c)	\\ 
2MASS J12531240+4034038	&	LHS2645		&	M7.5	&	$<$12			&	8.0 $\pm$ 2.0           	&	42	& (c)	\\ 
2MASS J15460540+3749458	&		-		&	M7.5	&	$<$12			&	10.0 $\pm$ 2.0          	&	35 	&	(c) \\ 	
2MASS J17571539+7042011	&	LP44-162		&	M7.5	&	39.2 $\pm$ 5.0		&	33.0 $\pm$ 3.0          	&	60 	&	(c)\\ 	
2MASS J00275592+2219328	&	LP349-25		&	M8.0	&	48.1 $\pm$ 6.0		&	56.0 $\pm$ 6.0          	&	104 	& (c)	\\ 	
2MASS J02532028+2713332	&		-		&	M8.0	&	17.5 $\pm$ 2.0		&	-			            	&	45	& 	\\ 
2MASS J03205965+1854233	&	LP412-31		&	M8.0	&	14.6 $\pm$ 3.0		&	15 $\pm$ 4.5            	&	65	& (c)	\\ 
2MASS J03510004-0052452	&	LHS1604		&	M8.0	&	$<$12			&	6.5 $\pm$ 2.0           	&	78 	&	(c)\\ 	
2MASS J05173766-3349027	&		-		&	M8.0	&	$<$12			&	8.0 $\pm$ 2.0           	&	48	& (c)	\\ 
2MASS J18432213+4040209	&	LHS3406		&	M8.0	&	$<$12			&	5.0 $\pm$ 3.0           	&	73	&	(c)\\ 	
2MASS J19165762+0509021	&	vB10			&	M8.0	&	template			&	-			            	&	80		& -	\\ 	
2MASS J22062280-2047058	&		-		&	M8.0	&	22.2 $\pm$ 2.0		&	24 $\pm$ 2.0            	&	44	& (c) \\ 	
2MASS J23494899+1224386	&	LP523-55		&	M8.0	&	$<$12			&	4 $\pm$ 2.0	            	&	29	& (c)	\\ 
2MASS J01400263+2701505	&		-		&	M8.5	&	$<$12			&	11			            	&	43	& (d)	\\ 
2MASS J18353790+3259545	&		-		&	M8.5	&	37.6 $\pm$ 5.0		&	44.0 $\pm$ 4.0          	&	119	& (c)	\\ 
2MASS J01095117-0343264	&	LP647-13		&	M9.0	&	$<$12			&	13 $\pm$ 2.0            	&	50	& (c)	\\ 
2MASS J03505737+1818069	&	LP413-53		&	M9.0	&	12.2 $\pm$ 3.0		&	-			            	&	34	&	\\ 
2MASS J04433761+0002051	&		-		&	M9.0	&	13.1 $\pm$ 2.0		&	13.5 $\pm$ 2.0          	&	42	& (c)	\\ 
2MASS J1733189+463359 	&		-		&	M9.5	&	18.2 $\pm$ 3.0		&	-			            	&	29	&	\\  

\enddata
\tablerefs{ v~sin~$i$ values from literature: (a) Mohanty $\&$ Basri (2003); (b) Delfosse et al. (1998); (c) Reiners $\&$ Basri, (2010); (d) Rice et al. (2010) }

\end{deluxetable}

\begin{deluxetable}{c c c c c c c c}
\rotate
\tablewidth{0pt}
\tablecaption{Radial Velocity (RV), Trigonometric Parallax, and Photometric Distances}
\tablehead{
\colhead{Object}           & \colhead{Alt.Name}      &
\colhead{Sp. Type}          & \colhead{RV (km/s)}  &
 \colhead{Lit.RV (km/s)} & 
\colhead{Trig. Para. (arcsec)}    & \colhead{Spectrophoto. (pc)}  &
\colhead{Ref.} 
} 
\tabletypesize
\scriptsize{}
\startdata
 2MASS J12185939+1107338 &	GJ1156		&	M5.0	&	7.7	$\pm$	1.4	&	5.86		  		&	0.1529 $\pm$ 0.0030	&	-				&	1,a	\\		
2MASS J16142520-0251009   &	LP624-54		&	M5.0	&	12.8	$\pm$	1.2	&	 -			    	&	-					&	14.60			&	2	\\
2MASS J22081254+1036420	&	RXJ2208.2	&	M5.0	&	-21.0$\pm$	1.3	&	-			    	&	-					&	44.00			&	3	\\
2MASS J23334057-2133526 	&	LHS3970		&	M5.0	&	27.9	$\pm$	1.2	&	-			  	&	-					&	24.10			&	4	\\
2MASS J00045753-1709369 	&	-			&	M5.5	&        3.2	$\pm$	3.8	&	-			    	&	-					&	15.9				&	4	\\
2MASS J00064325-0732147 	&	GJ1002		&	M5.5	&	-33.7$\pm$	3.8	&-35.7 $\pm$ 5.5    		&	0.2130 $\pm$ 0.0036	&	-				&	1,b	\\
2MASS J00130931-0025521 	&	-			&	M6.0	&	19.9	$\pm$	2.3	&	-			    	&	-					&	26.3				&	4	\\
2MASS J15524460-2623134 	&	LP860-41		&	M6.0	&	12.0	$\pm$	1.6	&	-			    	&	-					&	9.80				&	2	\\
2MASS J23415498+4410407 	&	GJ905		&	M6.0	&	-75.2$\pm$	3.7	&	-81			    	&	0.3160 $\pm$ 0.0011	&	-				&	1,c	\\
2MASS J02141251-0357434	&	LHS1363		&	M6.5	&	-15.6$\pm$	1.7	&	-			    	&	-					&	14.20 $\pm$ 1.50	&	5	\\
2MASS J03542008-1437388	&	-			&	M6.5	&	9.9	$\pm$	2.0	&	-			    	&	-					&	-				&	-	\\
2MASS J10562886+0700527	&	GJ406		&	M6.0	&	template   	1.5	&	19			    	&	0.4191 $\pm$ 0.0020	&	-				&	1,d	\\
2MASS J04351612-1606574	&	LP775-31		&	M7.0	&	48.5	$\pm$	1.4	&	52.5		    		&	-					&	11.30 $\pm$ 1.30	&	5,c	\\
2MASS J04402325-0530082	&	LP655-48		&	M7.0	&	31.1	$\pm$	1.4	&	27.5		    		&	-					&	9.80				&	6,c	\\
2MASS J07410681+1738459 	&	LHS1937		&	M7.0	&	41.8	$\pm$	1.2	&	38.6		    		&	-					&	17.90 $\pm$ 2.10	&	7,c	\\
2MASS J14563831-2809473	&	LHS3003		&	M7.0	&	1.0	$\pm$	1.5	&	0.9			    	&	0.1563 $\pm$ 0.0030	&	-				&	1,c	\\
2MASS J23312174-2749500	&	-			&	M7.0	&	-3.0	$\pm$	1.5	&	-			    	&	-					&	15.1				&	4	\\
2MASS J04173745-0800007 	&	-			&	M7.5	&	40.6 	$\pm$	1.4	&	38.4		    		&	-					&	17.40 $\pm$ 1.70	&	7,c	\\
2MASS J12531240+4034038	&	LHS2645		&	M7.5	&	4.3	 $\pm$	1.7	&	3.6			    	&	-					&	17.50 $\pm$ 1.70	&	7,c	\\
2MASS J15460540+3749458	&	-			&	M7.5	&	-22.1 $\pm$	2.5	&	-24.9		    	&	-					&	19.70 $\pm$ 1.90	&	7,c	\\
2MASS J17571539+7042011	&	LP44-162		&	M7.5	&	-15.3 $\pm$	2.8	&	-13.5		    	&	-					&	12.50 $\pm$ 1.20	&	7,c	\\
2MASS J00275592+2219328	&	LP349-25		&	M8.0	&	-15.8 $\pm$	1.9	&	-16.8		    	&	-					&	10.30 $\pm$ 1.70	&	7,c	\\
2MASS J02532028+2713332	&	-			&	M8.0	&	49.0 	$\pm$	1.5	&	-			    	&	-					&	18.5				&	8	\\
2MASS J03205965+1854233	&	LP412-31		&	M8.0	&	46.4	 $\pm$	1.3	&	44.9		    		&	0.0689 $\pm$ 0.0006	&	-				&	4,c	\\
2MASS J03510004-0052452	&	LHS1604		&	M8.0	&	-11.9 $\pm$	2.0	&	-14.7		    	&	0.0681 $\pm$ 0.0089	&	-				&	6,c	\\
2MASS J05173766-3349027	&	-			&	M8.0	&	29.4	$\pm$	2.8	&	31.4		    		&	16.4					&	-				&	6,c	\\
2MASS J18432213+4040209	&	LHS3406		&	M8.0	&	-19.3 $\pm$     2.0	&	-22.3		    	&	-					&	14.14 $\pm$ 0.16	&	9,c	\\
2MASS J19165762+0509021	&	vB10			&	M8.0	&	template            &	template	    			&	0.17					&	-				&	4	\\
2MASS J22062280-2047058	&	-			&	M8.0	&	10.8	$\pm$	1.6	&	9.8			    	&	-					&	18.2				&	4,c	\\
2MASS J23494899+1224386	&	LP523-55		&	M8.0	&	-3.7	$\pm$	1.9	&	-3.5		    		&	-					&	10.90 $\pm$ 0.60	&	5,c	\\
2MASS J01400263+2701505	&	-			&	M8.5	&	12.6	$\pm$	2.0	&	8.6			    	&	-					&	17.3				&	8,e	\\
2MASS J18353790+3259545	&	-			&	M8.5	&	8.4	$\pm$	2.0	&	8.5			    	&	-					&	-				&	7,c	\\
2MASS J01095117-0343264	&	LP647-13		&	M9.0	&	-6.3	$\pm$	1.8	&	-6.5		    		&	-					&	10.50			&	5	\\
2MASS J03505737+1818069	&	LP413-53		&	M9.0	&	32.2	$\pm$	1.8	&	-			    	&	-					&	20.30 $\pm$ 3.30	&	5	\\
2MASS J04433761+0002051	&	-			&	M9.0	&	21.9	$\pm$	2.3	&	17.1		    		&	-					&	16.20 $\pm$ 2.10	&	7,c	\\
2MASS J1733189+463359 	&	-			&	M9.5	&	-9.0	$\pm$	3.0	&	-			    	&	-					&	21.0 $\pm$ 1.30	&	7	\\
\enddata
\tablerefs{RV reference--:
(a) Morin et al. 2010; (b) Tinney et al. (1998); (c) Reiners et al. 2009; (d) Martin et al. (1997); (e) Rice et al. (2010). \textbf{References in column 6 $\&$ 7 are as follows}: \textbf{Trig. parallaxes and photometric distances:} (1)  chara.gsu.edu; (2) Phan Bao et al. (2008); (3) Zickgraf et al. (2005); (4) Crifo et al.(2005); (5) Cruz $\&$ Reid (2002); (6) Phan Bao et al. (2003); (7) Cruz et al. (2007); (8) Allen et al. (2007) }
\end{deluxetable}

\clearpage

\begin{deluxetable}{l c c c c c c}
\tablewidth{0pt}
\tablecaption{Pseudo-Equivalent Widths of K~I lines}
\tablehead{
\colhead{Object}           & \colhead{Alt.Name}      &
\colhead{Sp. Type}          & \colhead{11690 \AA $\ $}  &
\colhead{11771 \AA $\ $}          & \colhead{12435 \AA $\ $} &
\colhead{12522 \AA $\ $}
}
\tabletypesize
\scriptsize{}
\startdata
2MASSJ12185939+1107338	&	GJ1156	&	M5.0	&	1.98	$\pm$	0.15	&	2.82	$\pm$	0.10	&	1.45	$\pm$	0.06	&	1.69	$\pm$	0.08	\\
2MASSJ22081254+1036420	&	RXJ2208.2&	M5.0	&	1.76	$\pm$	0.08	&	2.30	$\pm$	0.18	&	1.20	$\pm$	0.13	&	1.32	$\pm$	0.13	\\
2MASSJ23334057-2133526	&	LHS3970	&	M5.0	&	1.97	$\pm$	0.08	&	3.08	$\pm$	0.10	&	1.56	$\pm$	0.07	&	1.65	$\pm$	0.07	\\
2MASSJ16142520-0251009	&	LP624-54	&	M5.0	&	2.68	$\pm$	0.13	&	4.25	$\pm$	0.15	&	1.90	$\pm$	0.11	&	2.17	$\pm$	0.12	\\
2MASSJ00045753-1709369	&	-		&	M5.5	&	2.28	$\pm$	0.13	&	3.47	$\pm$	0.15	&	1.94	$\pm$	0.12	&	2.07	$\pm$	0.12	\\
2MASSJ00064325-0732147	&	GJ1002	&	M5.5	&	2.25	$\pm$	0.16	&	3.26	$\pm$	0.41	&	1.95	$\pm$	0.08	&	2.09	$\pm$	0.32	\\
2MASSJ23415498+4410407	&	GJ905	&	M6.0	&	2.53	$\pm$	0.18	&	3.44	$\pm$	0.11	&	1.82	$\pm$	0.07	&	1.92	$\pm$	0.09	\\
2MASSJ15524460-2623134	&	LP860-41	&	M6.0	&	3.00	$\pm$	0.05	&	4.77	$\pm$	0.13	&	2.41	$\pm$	0.05	&	2.91	$\pm$	0.05	\\
2MASSJ00130931-0025521	&	-		&	M6.0	&	2.50	$\pm$	0.09	&	3.68	$\pm$	0.13	&	2.15	$\pm$	0.08	&	2.19	$\pm$	0.08	\\
2MASSJ10562886+0700527	&	GJ406	&	M6.0	&	3.01	$\pm$	0.05	&	4.81	$\pm$	0.08	&	2.41	$\pm$	0.12	&	2.89	$\pm$	0.06	\\
2MASSJ02141251-0357434	&	LHS1363	&	M6.5	&	3.11	$\pm$	0.10	&	4.31	$\pm$	0.11	&	2.32	$\pm$	0.08	&	2.57	$\pm$	0.07	\\
2MASSJ03542008-1437388	&	-		&	M6.5	&	3.10	$\pm$	0.16	&	5.46	$\pm$	0.41	&	2.37	$\pm$	0.14	&	3.24	$\pm$	0.33	\\
2MASSJ04351612-1606574	&	LP775-31	&	M7.0	&	4.48	$\pm$	0.16	&	6.66	$\pm$	0.39	&	3.62	$\pm$	0.28	&	4.34	$\pm$	0.31	\\
2MASSJ04402325-0530082	&	LP655-48	&	M7.0	&	4.36	$\pm$	0.19	&	6.75	$\pm$	0.23	&	3.35	$\pm$	0.17	&	3.99	$\pm$	0.18	\\
2MASSJ14563831-2809473	&	LHS3003	&	M7.0	&	4.19	$\pm$	0.31	&	6.40	$\pm$	0.20	&	3.32	$\pm$	0.15	&	3.97	$\pm$	0.16	\\
2MASSJ07410681+1738459	&	LHS1937	&	M7.0	&	4.15	$\pm$	0.33	&	6.40	$\pm$	0.20	&	3.40	$\pm$	0.29	&	3.49	$\pm$	0.16	\\
2MASSJ23312174-2749500	&	-		&	M7.0	&	4.34	$\pm$	0.27	&	7.12	$\pm$	0.22	&	3.64	$\pm$	0.24	&	5.02	$\pm$	0.25	\\
2MASSJ17571539+7042011	&	LP44-162	&	M7.5	&	4.21	$\pm$	0.31	&	6.29	$\pm$	0.37	&	3.45	$\pm$	0.27	&	3.73	$\pm$	0.30	\\
2MASSJ12531240+4034038	&	LHS2645	&	M7.5	&	4.15	$\pm$	0.27	&	5.68	$\pm$	0.37	&	3.28	$\pm$	0.24	&	3.31	$\pm$	0.28	\\
2MASSJ15460540+3749458	&	-		&	M7.5	&	4.34	$\pm$	0.43	&	6.02	$\pm$	0.48	&	3.34	$\pm$	0.39	&	3.43	$\pm$	0.42	\\
2MASSJ04173745-0800007	&	-		&	M7.5	&	4.42	$\pm$	0.12	&	6.78	$\pm$	0.41	&	3.61	$\pm$	0.22	&	4.16	$\pm$	0.38	\\
2MASSJ19165762+0509021	&	vB10	&	M8.0	&	5.05	$\pm$	0.63	&	7.09	$\pm$	0.48	&	3.58	$\pm$	0.24	&	4.37	$\pm$	0.32	\\
2MASSJ23494899+1224386	&	LP523-55	&	M8.0	&	5.08	$\pm$	0.28	&	7.06	$\pm$	0.25	&	3.83	$\pm$	0.18	&	4.20	$\pm$	0.39	\\
2MASSJ03205965+1854233	&	LP412-31	&	M8.0	&	5.30	$\pm$	0.17	&	7.64	$\pm$	0.19	&	3.88	$\pm$	0.20	&	5.10	$\pm$	0.15	\\
2MASSJ00275592+2219328	&	LP349-25	&	M8.0	&	4.43	$\pm$	0.17	&	6.12	$\pm$	0.29	&	3.36	$\pm$	0.13	&	3.58	$\pm$	0.14	\\
2MASSJ18432213+4040209	&	LHS3406	&	M8.0	&	4.44	$\pm$	0.34	&	6.21	$\pm$	0.14	&	3.39	$\pm$	0.34	&	3.69	$\pm$	0.58	\\
2MASSJ03510004-0052452	&	LHS1604	&	M8.0	&	4.43	$\pm$	0.34	&	6.04	$\pm$	0.31	&	3.21	$\pm$	0.29	&	3.38	$\pm$	0.32	\\
2MASSJ05173766-3349027	&	-		&	M8.0	&	5.07	$\pm$	0.26	&	7.26	$\pm$	0.34	&	3.63	$\pm$	0.23	&	5.27	$\pm$	0.24	\\
2MASSJ22062280-2047058	&	-		&	M8.0	&	5.01	$\pm$	0.19	&	7.12	$\pm$	0.22	&	3.63	$\pm$	0.17	&	4.67	$\pm$	0.18	\\
2MASSJ02532028+2713332	&	-		&	M8.0	&	5.07	$\pm$	0.45	&	7.44	$\pm$	0.30	&	3.73	$\pm$	0.22	&	4.59	$\pm$	0.23	\\
2MASSJ18353790+3259545	&	-		&	M8.5	&	5.49	$\pm$	0.24	&	7.94	$\pm$	0.20	&	4.06	$\pm$	0.19	&	5.05	$\pm$	0.17	\\
2MASSJ01400263+2701505	&	-		&	M8.5	&	5.97	$\pm$	0.41	&	7.06	$\pm$	0.32	&	4.45	$\pm$	0.31	&	4.88	$\pm$	0.48	\\
2MASSJ01095117-0343264	&	LP647-13	&	M9.0	&	5.96	$\pm$	0.30	&	7.61	$\pm$	0.47	&	3.65	$\pm$	0.36	&	4.54	$\pm$	0.34	\\
2MASSJ03505737+1818069	&	LP413-53	&	M9.0	&	5.83	$\pm$	0.37	&	6.58	$\pm$	0.38	&	3.94	$\pm$	0.25	&	3.97	$\pm$	0.25	\\
2MASSJ04433761+0002051	&	-		&	M9.0	&	5.90	$\pm$	0.58	&	6.09	$\pm$	0.59	&	3.61	$\pm$	0.47	&	3.50	$\pm$	0.48	\\
2MASSJ1733189+463359	&	-		&	M9.5	&	6.17	$\pm$	0.48	&	7.67	$\pm$	0.53	&	4.42	$\pm$	0.41	&	4.91	$\pm$	0.44	\\																				
\enddata
\end{deluxetable}

\clearpage

\begin{deluxetable}{l c c c c c c}
\tablewidth{0pt}
\tablecaption{Pseudo-Equivalent Widths of Fe~I $\&$ Na I}
\tablehead{
\colhead{Object}           & \colhead{Alt.Name}      &
\colhead{Sp. Type}          & \colhead{11786 \AA $\ $}  &
\colhead{11886 \AA $\ $}          & \colhead{11976 \AA $\ $} &
\colhead{12682 \AA $\ $}
}
\tabletypesize
\scriptsize{}
\startdata
2MASSJ12185939+1107338	&	GJ1156		&	M5.0   &	0.22	$\pm$	0.01	&	1.14	$\pm$	0.05	&	1.04	$\pm$	0.05	&	0.52	$\pm$	0.01	\\
2MASSJ22081254+1036420	&	RXJ2208.2	&	M5.0   &	0.26	$\pm$	0.02	&	1.31	$\pm$	0.09	&	1.10	$\pm$	0.09	&	0.52	$\pm$	0.02	\\
2MASSJ23334057-2133526	&	LHS3970		&	M5.0   &	0.23	$\pm$	0.02	&	1.15	$\pm$	0.05	&	1.17	$\pm$	0.05	&	0.51	$\pm$	0.04	\\
2MASSJ16142520-0251009	&	LP624-54	&	M5.0   &	0.25	$\pm$	0.04	&	1.21	$\pm$	0.10	&	1.27	$\pm$	0.09	&	0.62	$\pm$	0.03	\\
2MASSJ00045753-1709369	&	-			&	M5.5   &	0.26	$\pm$	0.01	&	1.24	$\pm$	0.08	&	1.09	$\pm$	0.07	&	0.64	$\pm$	0.04	\\
2MASSJ00064325-0732147	&	GJ1002		&	M5.5   &	0.22	$\pm$	0.01	&	1.11	$\pm$	0.20	&	1.09	$\pm$	0.19	&	0.66	$\pm$	0.03	\\
2MASSJ23415498+4410407	&	GJ905		&	M6.0   &	0.18	$\pm$	0.01	&	0.77	$\pm$	0.04	&	0.95	$\pm$	0.05	&	0.73	$\pm$	0.02	\\
2MASSJ15524460-2623134	&	LP860-41	&	M6.0   &	0.25	$\pm$	0.03	&	1.03	$\pm$	0.03	&	1.26	$\pm$	0.03	&	0.65	$\pm$	0.03	\\
2MASSJ00130931-0025521	&	-			&	M6.0   &	0.26	$\pm$	0.03	&	1.23	$\pm$	0.03	&	1.12	$\pm$	0.03	&	0.49	$\pm$	0.01	\\
2MASSJ10562886+0700527	&	GJ406		&	M6.0   &	0.22	$\pm$	0.01	&	1.08	$\pm$	0.03	&	1.17	$\pm$	0.03	&	0.66	$\pm$	0.02	\\
2MASSJ02141251-0357434	&	LHS1363		&	M6.5   &	0.23	$\pm$	0.02	&	1.16	$\pm$	0.05	&	1.36	$\pm$	0.05	&	0.59	$\pm$	0.03	\\
2MASSJ03542008-1437388	&	-			&	M6.5   &	0.25	$\pm$	0.04	&	1.42	$\pm$	0.16	&	1.34	$\pm$	0.16	&	0.49	$\pm$	0.04	\\
2MASSJ04351612-1606574	&	LP775-31	&	M7.0   &	0.28	$\pm$	0.04	&	1.27	$\pm$	0.13	&	1.51	$\pm$	0.15	&	\nodata			\\
2MASSJ04402325-0530082	&	LP655-48	&	M7.0   &	0.28	$\pm$	0.05	&	1.21	$\pm$	0.08	&	1.50	$\pm$	0.09	&	0.51	$\pm$	0.03	\\
2MASSJ14563831-2809473	&	LHS3003		&	M7.0   &	0.27	$\pm$	0.03	&	1.30	$\pm$	0.07	&	1.68	$\pm$	0.08	&	0.51	$\pm$	0.05	\\
2MASSJ07410681+1738459	&	LHS1937		&	M7.0   &	0.24	$\pm$	0.04	&	1.14	$\pm$	0.07	&	1.51	$\pm$	0.08	&	0.54	$\pm$	0.07	\\
2MASSJ23312174-2749500	&	-			&	M7.0   &	0.24	$\pm$	0.04	&	1.44	$\pm$	0.11	&	1.66	$\pm$	0.12	&	\nodata			\\
2MASSJ17571539+7042011	&	LP44-162	&	M7.5   &	\nodata					&	1.39	$\pm$	0.14	&	1.20	$\pm$	0.14	&	\nodata			\\
2MASSJ12531240+4034038	&	LHS2645		&	M7.5   &	0.25	$\pm$	0.06	&	1.28	$\pm$	0.13	&	1.33	$\pm$	0.12	&	0.49	$\pm$	0.05	\\
2MASSJ15460540+3749458	&	-			&	M7.5   &	\nodata					&	1.41	$\pm$	0.20	&	1.38	$\pm$	0.20	&	0.52	$\pm$	0.04	\\
2MASSJ04173745-0800007	&	-			&	M7.5   &	0.24	$\pm$	0.08	&	1.36	$\pm$	0.15	&	1.34	$\pm$	0.17	&	0.53	$\pm$	0.03	\\
2MASSJ19165762+0509021	&	vB10		&	M8.0   &	\nodata					&	1.19	$\pm$	0.14	&	1.64	$\pm$	0.15	&	0.48	$\pm$	0.05	\\
2MASSJ23494899+1224386	&	LP523-55	&	M8.0   &	\nodata					&	0.86	$\pm$	0.14	&	1.60	$\pm$	0.18	&	0.29	$\pm$	0.05	\\
2MASSJ03205965+1854233	&	LP412-31	&	M8.0   &	\nodata					&	1.45	$\pm$	0.07	&	1.51	$\pm$	0.07	&	0.36	$\pm$	0.03	\\
2MASSJ00275592+2219328	&	LP349-25	&	M8.0   &	\nodata					&	1.16	$\pm$	0.06	&	1.16	$\pm$	0.07	&	0.26	$\pm$	0.02	\\
2MASSJ18432213+4040209	&	LHS3406		&	M8.0   &	\nodata					&	0.98	$\pm$	0.21	&	1.75	$\pm$	0.28	&	0.49	$\pm$	0.12	\\
2MASSJ03510004-0052452	&	LHS1604		&	M8.0   &	\nodata					&	1.09	$\pm$	0.14	&	1.60	$\pm$	0.16	&	\nodata			\\
2MASSJ05173766-3349027	&	-			&	M8.0   &	\nodata					&	0.93	$\pm$	0.11	&	1.65	$\pm$	0.11	&	\nodata			\\
2MASSJ22062280-2047058	&	-			&	M8.0   &	\nodata					&	1.18	$\pm$	0.08	&	1.57	$\pm$	0.08	&	0.36	$\pm$	0.05	\\
2MASSJ02532028+2713332	&	-			&	M8.0   &	\nodata					&	1.09	$\pm$	0.09	&	1.44	$\pm$	0.10	&	\nodata			\\
2MASSJ18353790+3259545	&	-			&	M8.5   &	\nodata					&	1.17	$\pm$	0.07	&	1.35	$\pm$	0.07	&	\nodata			\\
2MASSJ01400263+2701505	&	-			&	M8.5   &	\nodata					&	1.23	$\pm$	0.20	&	1.35	$\pm$	0.20	&	\nodata			\\
2MASSJ01095117-0343264	&	LP647-13	&	M9.0   &	\nodata					&	1.30	$\pm$	0.21	&	1.21	$\pm$	0.20	&	\nodata			\\
2MASSJ03505737+1818069	&	LP413-53	&	M9.0   &	\nodata					&	1.23	$\pm$	0.13	&	1.19	$\pm$	0.09	&	\nodata			\\
2MASSJ04433761+0002051	&	-			&	M9.0   &	\nodata					&	1.23	$\pm$	0.21	&	1.25	$\pm$	0.21	&	\nodata			\\
2MASSJ1733189+463359	&	-			&	M9.5   &	1.02	$\pm$	0.11	&	1.16	$\pm$	0.17	&	1.20	$\pm$	0.12	&	\nodata			\\

\enddata
\end{deluxetable}

\clearpage
	
\begin{deluxetable}{l c c c c c c c}
\tablewidth{0pt}
\tablecaption{Pseudo-Equivalent Widths of Ti I, Mn I, and Al I}
\tablehead{
\colhead{Object}           & \colhead{Alt.Name}      &
\colhead{Sp. Type}          & \colhead{12832 \AA $\ $}  &
\colhead{12850 \AA $\ $}          & \colhead{12899 \AA $\ $} &
\colhead{13123 \AA $\ $}       & \colhead{13150 \AA $\ $}
}
\tabletypesize
\scriptsize{}
\startdata
2MASSJ12185939+1107338	&	GJ1156	&	M5.0	&	0.13	$\pm$	0.01	&	0.22	$\pm$	0.01	&	0.49	$\pm$	0.02	&	1.15	$\pm$	0.04	&	0.72	$\pm$	0.03	\\
2MASSJ22081254+1036420	&	RXJ2208.2&	M5.0	&	0.13	$\pm$	0.02	&	0.24	$\pm$	0.02	&	0.56	$\pm$	0.04	&	1.36	$\pm$	0.08	&	0.96	$\pm$	0.07	\\
2MASSJ23334057-2133526	&	LHS3970	&	M5.0	&	0.13	$\pm$	0.02	&	0.23	$\pm$	0.02	&	0.47	$\pm$	0.02	&	1.22	$\pm$	0.04	&	0.76	$\pm$	0.04	\\
2MASSJ16142520-0251009	&	LP624-54	&	M5.0	&	0.16	$\pm$	0.05	&	0.22	$\pm$	0.03	&	0.48	$\pm$	0.04	&	1.25	$\pm$	0.09	&	0.84	$\pm$	0.07	\\
2MASSJ00045753-1709369	&	-		&	M5.5	&	0.24	$\pm$	0.01	&	0.17	$\pm$	0.02	&	0.46	$\pm$	0.03	&	1.25	$\pm$	0.07	&	0.79	$\pm$	0.05	\\
2MASSJ00064325-0732147	&	GJ1002	&	M5.5	&	0.24	$\pm$	0.01	&	0.19	$\pm$	0.05	&	0.47	$\pm$	0.09	&	1.34	$\pm$	0.17	&	0.90	$\pm$	0.13	\\
2MASSJ23415498+4410407	&	GJ905	&	M6.0	&	0.21	$\pm$	0.01	&	0.18	$\pm$	0.01	&	0.38	$\pm$	0.02	&	1.15	$\pm$	0.05	&	0.66	$\pm$	0.04	\\
2MASSJ15524460-2623134	&	LP860-41	&	M6.0	&	0.21	$\pm$	0.03	&	0.19	$\pm$	0.01	&	0.55	$\pm$	0.01	&	1.27	$\pm$	0.02	&	0.86	$\pm$	0.02	\\
2MASSJ00130931-0025521	&	-		&	M6.0	&	0.17	$\pm$	0.03	&	0.17	$\pm$	0.01	&	0.54	$\pm$	0.02	&	1.14	$\pm$	0.05	&	0.71	$\pm$	0.04	\\
2MASSJ10562886+0700527	&	GJ406	&	M6.0	&	0.20	$\pm$	0.01	&	0.20	$\pm$	0.01	&	0.55	$\pm$	0.01	&	1.19	$\pm$	0.03	&	0.68	$\pm$	0.02	\\
2MASSJ02141251-0357434	&	LHS1363	&	M6.5	&	0.20	$\pm$	0.03	&	0.20	$\pm$	0.01	&	0.57	$\pm$	0.02	&	1.25	$\pm$	0.03	&	0.73	$\pm$	0.02	\\
2MASSJ03542008-1437388	&	-		&	M6.5	&	0.17	$\pm$	0.04	&	0.19	$\pm$	0.03	&	0.57	$\pm$	0.07	&	1.17	$\pm$	0.13	&	0.71	$\pm$	0.08	\\
2MASSJ04351612-1606574	&	LP775-31	&	M7.0	&	0.15	$\pm$	0.04	&	0.20	$\pm$	0.03	&	0.49	$\pm$	0.06	&	1.25	$\pm$	0.12	&	0.58	$\pm$	0.08	\\
2MASSJ04402325-0530082	&	LP655-48	&	M7.0	&	0.15	$\pm$	0.03	&	0.18	$\pm$	0.02	&	0.59	$\pm$	0.04	&	1.29	$\pm$	0.07	&	0.66	$\pm$	0.05	\\
2MASSJ14563831-2809473	&	LHS3003	&	M7.0	&	0.15	$\pm$	0.03	&	0.20	$\pm$	0.02	&	0.59	$\pm$	0.03	&	1.26	$\pm$	0.06	&	0.66	$\pm$	0.04	\\
2MASSJ07410681+1738459	&	LHS1937	&	M7.0	&	0.16	$\pm$	0.03	&	0.22	$\pm$	0.02	&	0.55	$\pm$	0.03	&	1.26	$\pm$	0.06	&	0.63	$\pm$	0.04	\\
2MASSJ23312174-2749500	&	-		&	M7.0	&	0.16	$\pm$	0.02	&	0.20	$\pm$	0.03	&	0.55	$\pm$	0.05	&	1.29	$\pm$	0.09	&	0.55	$\pm$	0.06	\\
2MASSJ17571539+7042011	&	LP44-162	&	M7.5	&	0.14	$\pm$	0.02	&	\nodata			&	0.43	$\pm$	0.06	&	1.20	$\pm$	0.11	&	0.64	$\pm$	0.09	\\
2MASSJ12531240+4034038	&	LHS2645	&	M7.5	&	0.16	$\pm$	0.02	&	0.15	$\pm$	0.03	&	0.58	$\pm$	0.06	&	1.33	$\pm$	0.11	&	0.65	$\pm$	0.08	\\
2MASSJ15460540+3749458	&	-		&	M7.5	&	0.19	$\pm$	0.01	&	0.15	$\pm$	0.04	&	0.52	$\pm$	0.08	&	1.15	$\pm$	0.16	&	0.64	$\pm$	0.12	\\
2MASSJ04173745-0800007	&	-		&	M7.5	&	0.16	$\pm$	0.04	&	0.13	$\pm$	0.03	&	0.57	$\pm$	0.08	&	1.23	$\pm$	0.13	&	0.68	$\pm$	0.09	\\
2MASSJ19165762+0509021	&	vB10	&	M8.0	&	0.14	$\pm$	0.01	&	0.17	$\pm$	0.03	&	0.57	$\pm$	0.07	&	1.28	$\pm$	0.12	&	0.55	$\pm$	0.09	\\
2MASSJ23494899+1224386	&	LP523-55	&	M8.0	&	\nodata			&	0.18	$\pm$	0.03	&	0.69	$\pm$	0.07	&	1.12	$\pm$	0.12	&	0.65	$\pm$	0.09	\\
2MASSJ03205965+1854233	&	LP412-31	&	M8.0	&	0.15	$\pm$	0.03	&	0.14	$\pm$	0.01	&	0.59	$\pm$	0.04	&	1.25	$\pm$	0.06	&	0.64	$\pm$	0.04	\\
2MASSJ00275592+2219328	&	LP349-25	&	M8.0	&	\nodata			&	\nodata			&	0.46	$\pm$	0.03	&	0.99	$\pm$	0.05	&	0.50	$\pm$	0.04	\\
2MASSJ18432213+4040209	&	LHS3406	&	M8.0	&	0.16	$\pm$	0.04	&	0.16	$\pm$	0.06	&	0.61	$\pm$	0.13	&	1.41	$\pm$	0.20	&	0.67	$\pm$	0.16	\\
2MASSJ03510004-0052452	&	LHS1604	&	M8.0	&	\nodata			&	0.17	$\pm$	0.03	&	0.61	$\pm$	0.07	&	1.19	$\pm$	0.12	&	0.66	$\pm$	0.08	\\
2MASSJ05173766-3349027	&	-		&	M8.0	&	0.12	$\pm$	0.03	&	0.12	$\pm$	0.05	&	0.54	$\pm$	0.05	&	0.97	$\pm$	0.09	&	0.53	$\pm$	0.06	\\
2MASSJ22062280-2047058	&	-		&	M8.0	&	0.12	$\pm$	0.02	&	0.15	$\pm$	0.02	&	0.57	$\pm$	0.04	&	1.17	$\pm$	0.06	&	0.61	$\pm$	0.05	\\
2MASSJ02532028+2713332	&	-		&	M8.0	&	\nodata			&	0.15	$\pm$	0.04	&	0.57	$\pm$	0.04	&	1.19	$\pm$	0.07	&	0.49	$\pm$	0.10	\\
2MASSJ18353790+3259545	&	-		&	M8.5	&	0.07	$\pm$	0.01	&	\nodata			&	0.50	$\pm$	0.03	&	1.01	$\pm$	0.04	&	\nodata			\\
2MASSJ01400263+2701505	&	-		&	M8.5	&	0.17	$\pm$	0.02	&	0.18	$\pm$	0.04	&	0.54	$\pm$	0.10	&	1.27	$\pm$	0.10	&	0.64	$\pm$	0.10	\\
2MASSJ01095117-0343264	&	LP647-13	&	M9.0	&	0.14	$\pm$	0.03	&	0.11	$\pm$	0.03	&	0.60	$\pm$	0.08	&	1.02	$\pm$	0.13	&	0.56	$\pm$	0.10	\\
2MASSJ03505737+1818069	&	LP413-53	&	M9.0	&	0.16	$\pm$	0.04	&	0.13	$\pm$	0.03	&	0.57	$\pm$	0.04	&	1.13	$\pm$	0.10	&	0.67	$\pm$	0.08	\\
2MASSJ04433761+0002051	&	-		&	M9.0	&	0.18	$\pm$	0.09	&	0.13	$\pm$	0.04	&	0.44	$\pm$	0.10	&	0.99	$\pm$	0.17	&	0.50	$\pm$	0.13	\\
2MASSJ1733189+463359	&	-		&	M9.5	&	0.10	$\pm$	0.09	&	0.11	$\pm$	0.04	&	0.56	$\pm$	0.08	&	1.01	$\pm$	0.13	&	0.30	$\pm$	0.10	\\
\enddata
\end{deluxetable}

\clearpage

\begin{table}
\begin{center}
\caption{Relations between p-EW and SpT }
\begin{tabular}{ c  c  c  c  } 
\tableline\tableline
Wavelength		& 	\multicolumn{3}{c}{ y = mx+b}	 \\ [-0.05ex]
\cline{3-4}
	$\AA $\ $$		&      Element		&	m ($\AA $\ $$/SpT)		& 		b ($\AA $\ $$)      		 \\
\hline
11690			&	K~I		&	0.992  $\pm$ 0.062		&   -3.026 $\pm$ 0.440	 \\
11771			&	K~I 		&	1.265 $\pm$ 0.047		&   -3.256 $\pm$ 0.326	 \\
11886			&	Fe~I		&	0.033 $\pm$ 0.022		&   0.919 $\pm$ 0.075	 \\
11976			&	Fe~I		&	0.094  $\pm$ 0.026		&   0.634 $\pm$ 0.090	 \\
12435			&	K~I 		&	0.698  $\pm$ 0.084		&  -1.957 $\pm$ 0.599		\\
12522			&	K~I 		&	0.856  $\pm$ 0.064		&  -2.538 $\pm$ 0.447	 \\
12682			&	Na I		&	-0.077 $\pm$ 0.020		&   1.208 $\pm$ 0.064		\\
12832			&	Ti~I		&	-0.021 $\pm$ 0.001		&   0.321 $\pm$ 0.017		\\
12850			&	Ti~I		&	-0.021 $\pm$ 0.003		&   0.314 $\pm$ 0.010		\\
12900			&	Mn~I		&	0.015  $\pm$ 0.010		&   0.435 $\pm$ 0.030		\\
13127			& 	Al I		&	-0.039 $\pm$ 0.014		&   1.461 $\pm$ 0.047		\\
13150			&	Al I		&	-0.054 $\pm$ 0.010 		&   0.850 $\pm$ 0.034		\\
   											
\tableline
\end{tabular}
\end{center}
\end{table}

\clearpage




\begin{figure}
\begin{center}
\includegraphics[width=12.74 cm]{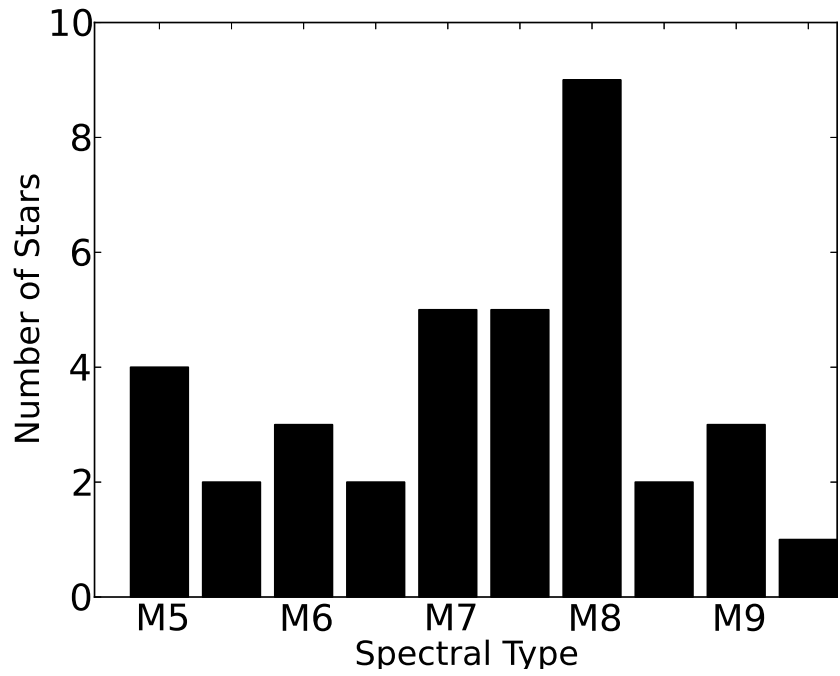} 
\end{center}
\caption{Distribution of our targets by sub-spectral types. The emphasis of late-M dwarfs in this study is visible through a large number of M8.0 targets.}
\end{figure}

\clearpage
\begin{figure*}
\begin{center} $
\begin{array}{lll}
\includegraphics[width=8.0cm]{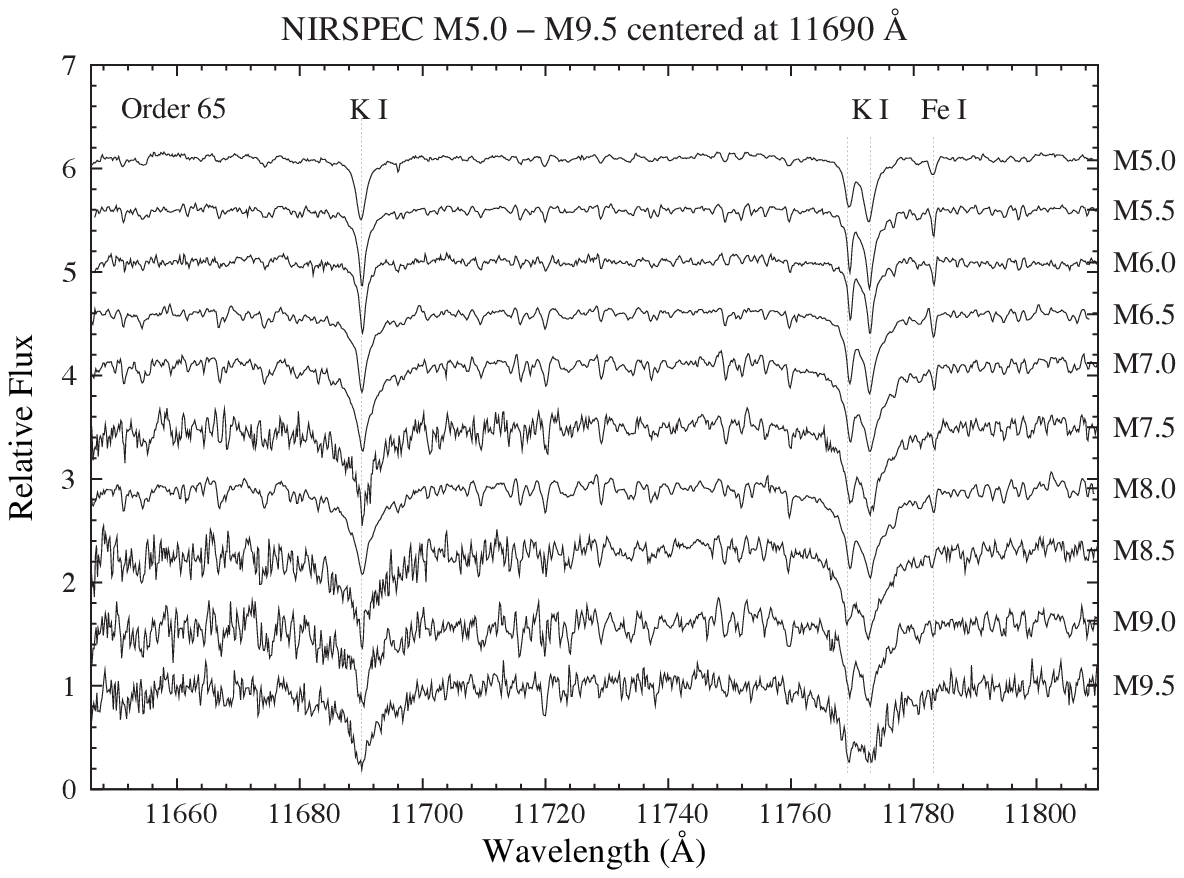}&
\includegraphics[width=8.0cm]{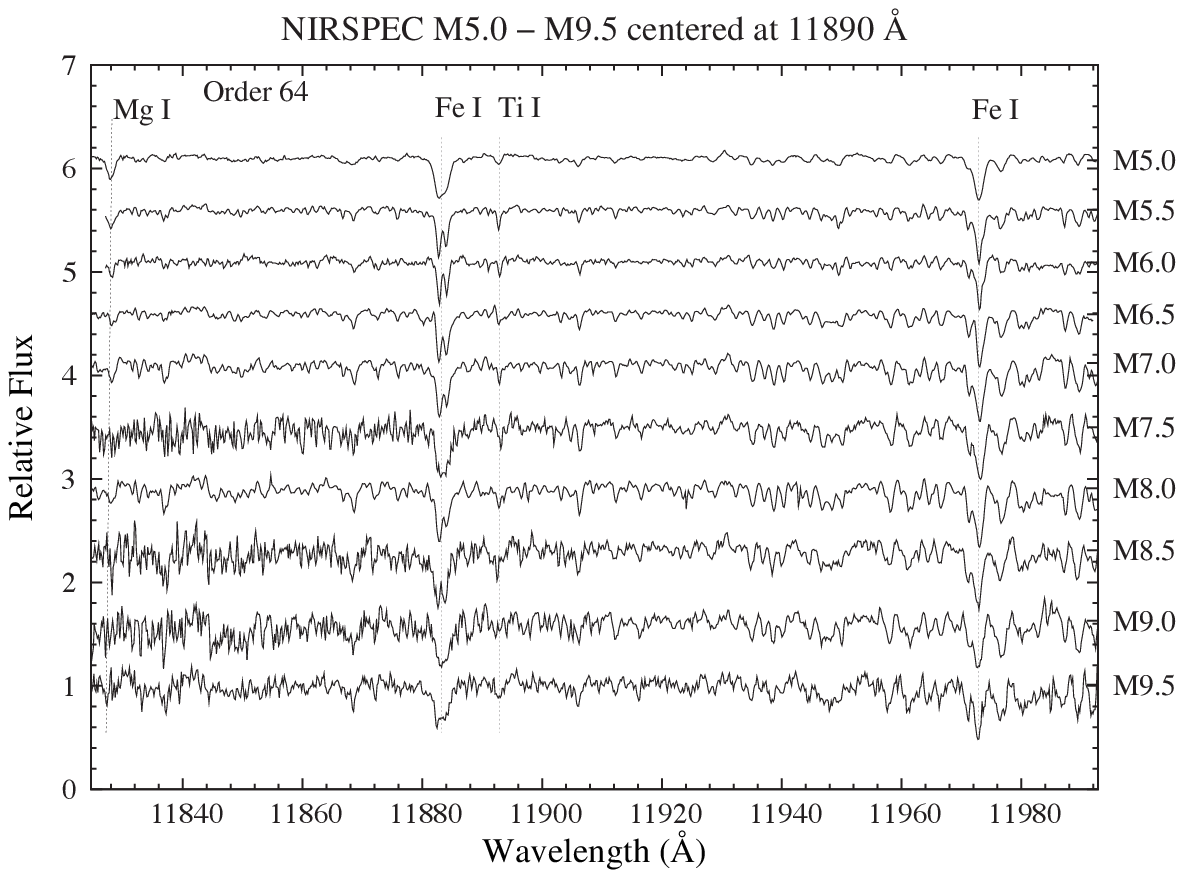}& \\
\includegraphics[width=8.0cm]{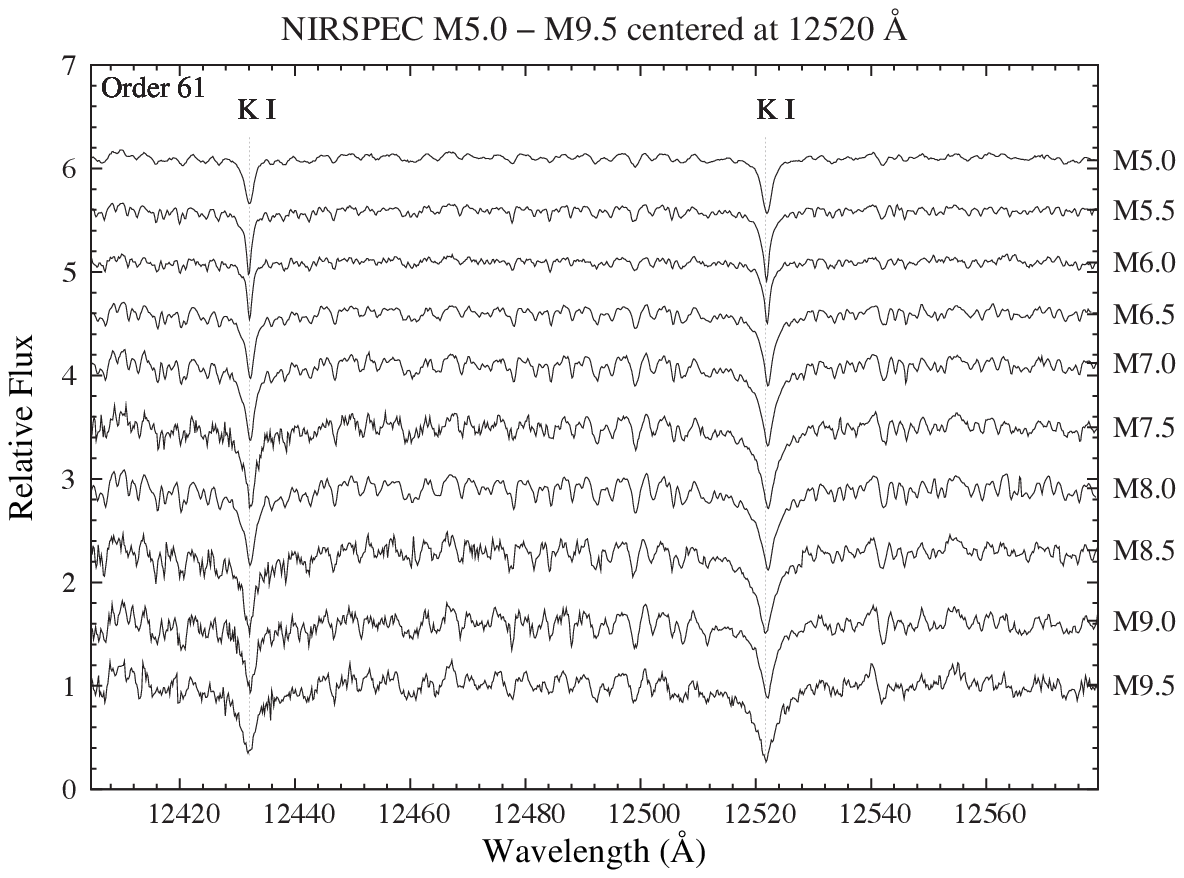}&
\includegraphics[width=8.0cm]{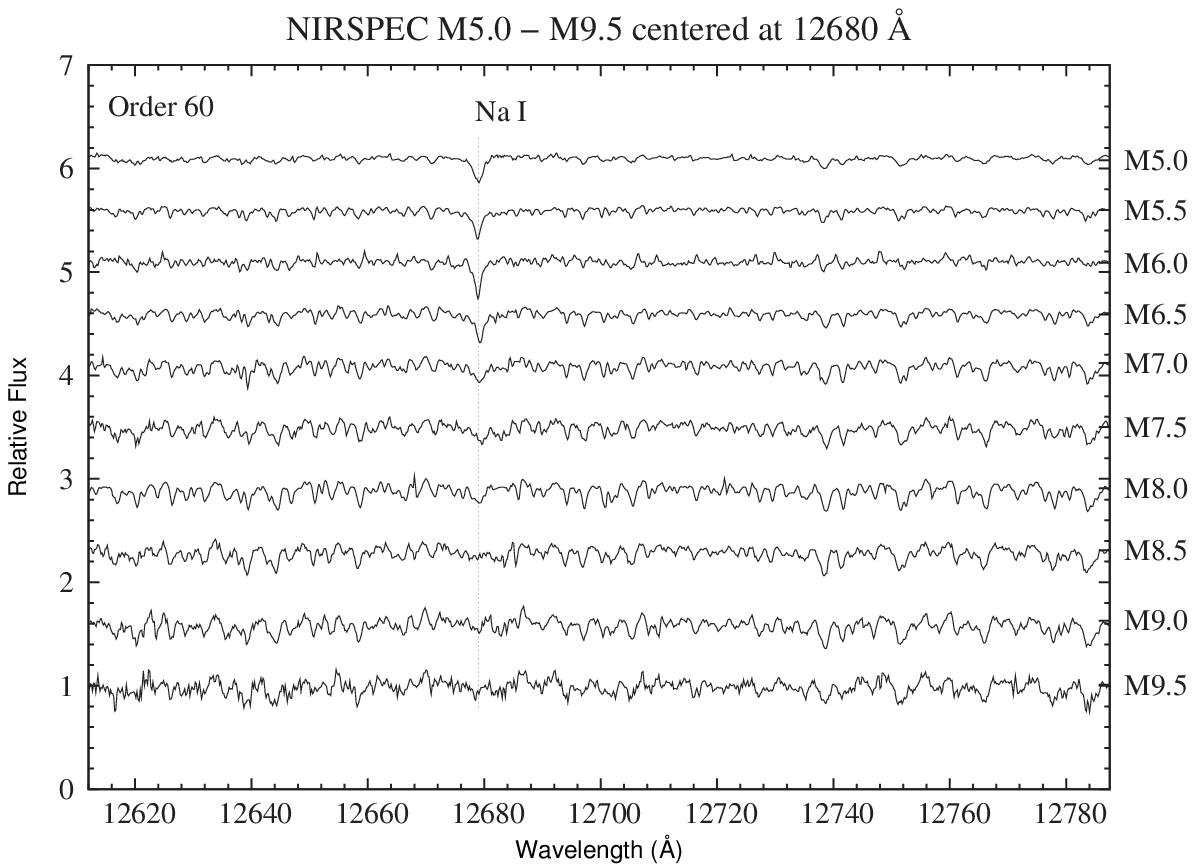} \\
\includegraphics[width=8.0cm]{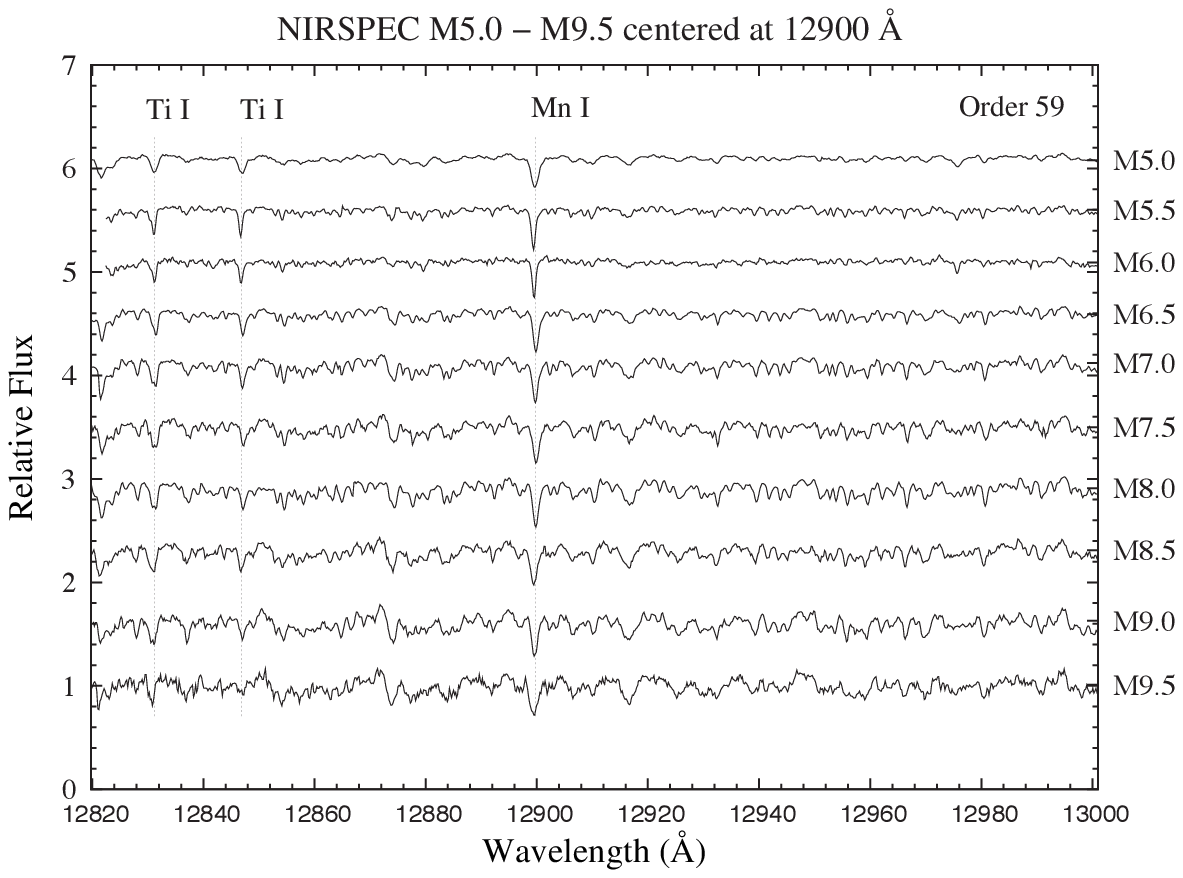}&
\includegraphics[width=8.0cm]{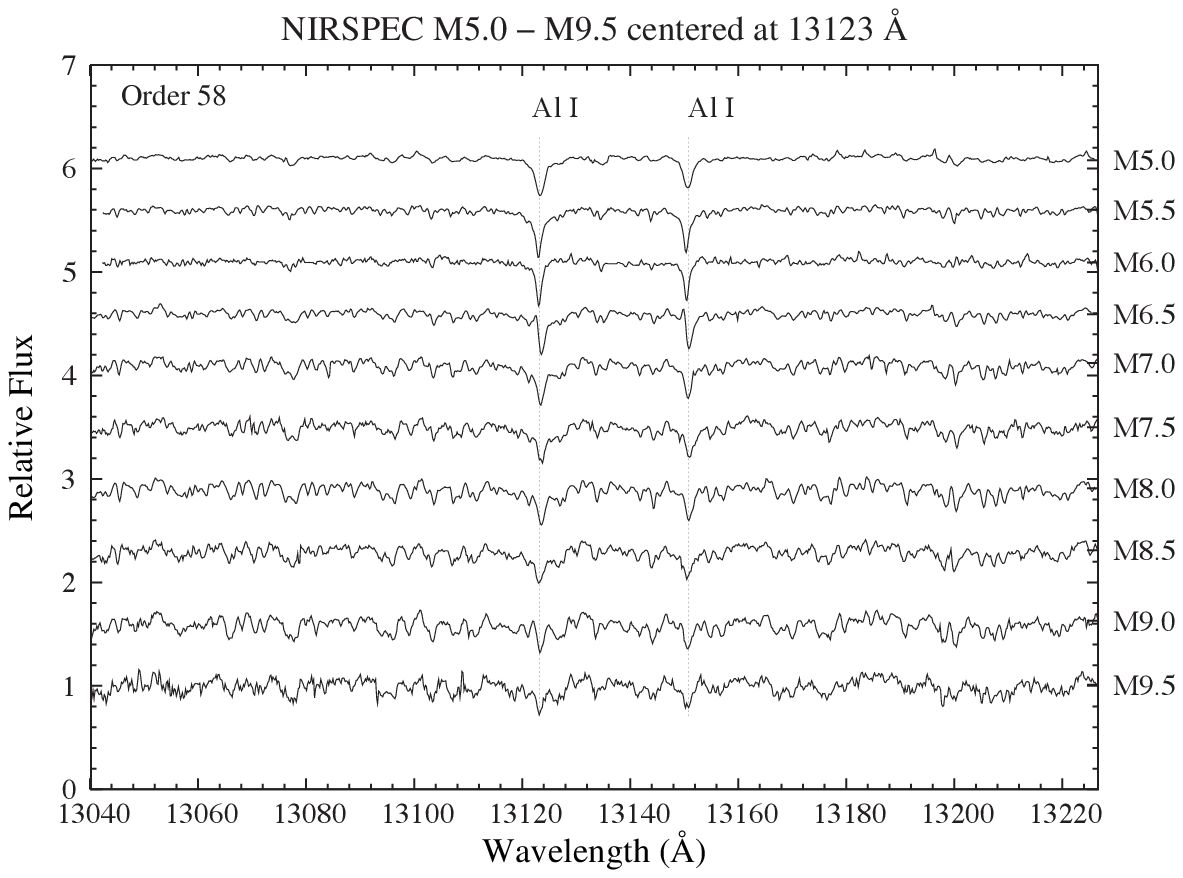}&
\end{array} $
\end{center}
\caption{Spectral sequence of our sample. The targets are GJ1156 (M5.0), GJ1002 (M5.5), GJ905 (M6.0), LHS1363 (M6.5), LHS3003 (M7.0), 2M0417-0800 (M7.5), vB10 (M8.0), 2M0140+2701 (M8.5), 2M0443+0002 (M9.0), 2M1733+4633 (M9.5). Each panel corresponds to a NIRSPEC echelle order. The spectra are normalized to unity and are offset by a constant. Each spectrum is shifted and centered at a specified wavelength. Each spectra is identified by its target. The absorption lines are indicated by the dashed lines. }
\end{figure*}


\begin{figure}
\begin{center}
\includegraphics[width=12.74 cm]{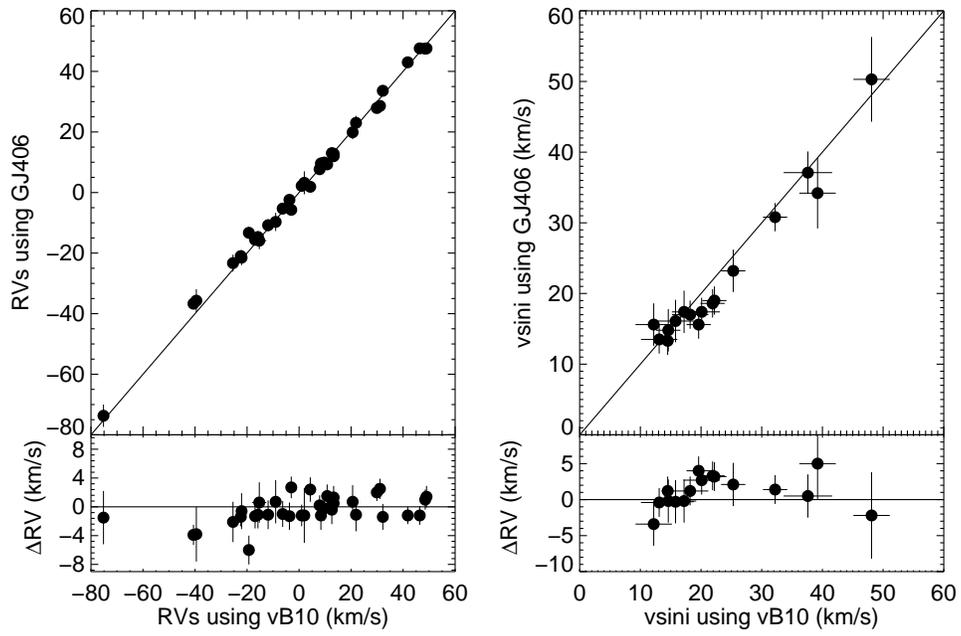} 
\end{center}
\caption{Comparison of absolute radial velocities (left panel), projected rotational velocities (right panel) and their associated residuals (bottom panels) using GJ406 and vB10 as templates.}
\end{figure}
\clearpage
\begin{figure}
\begin{center}
\includegraphics[width=12.74 cm]{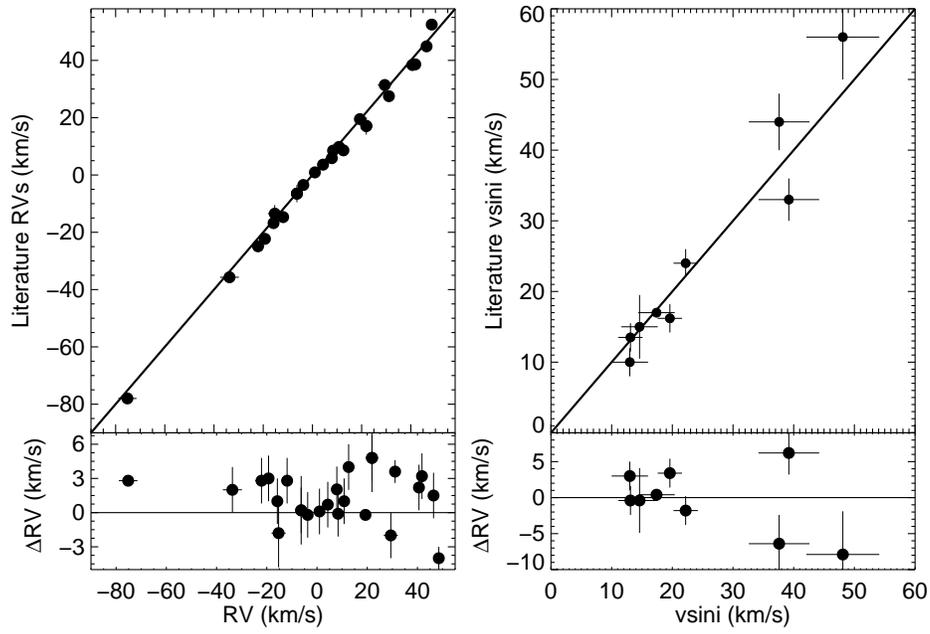} 
\end{center}
\caption{Comparison of absolute radial velocities (right panel), projected rotational velocities (left panel) and their associated residuals (bottom panels) with those in the literature.}
\end{figure}
\clearpage

\begin{figure}
\begin{center} 
\includegraphics[width=12.74 cm]{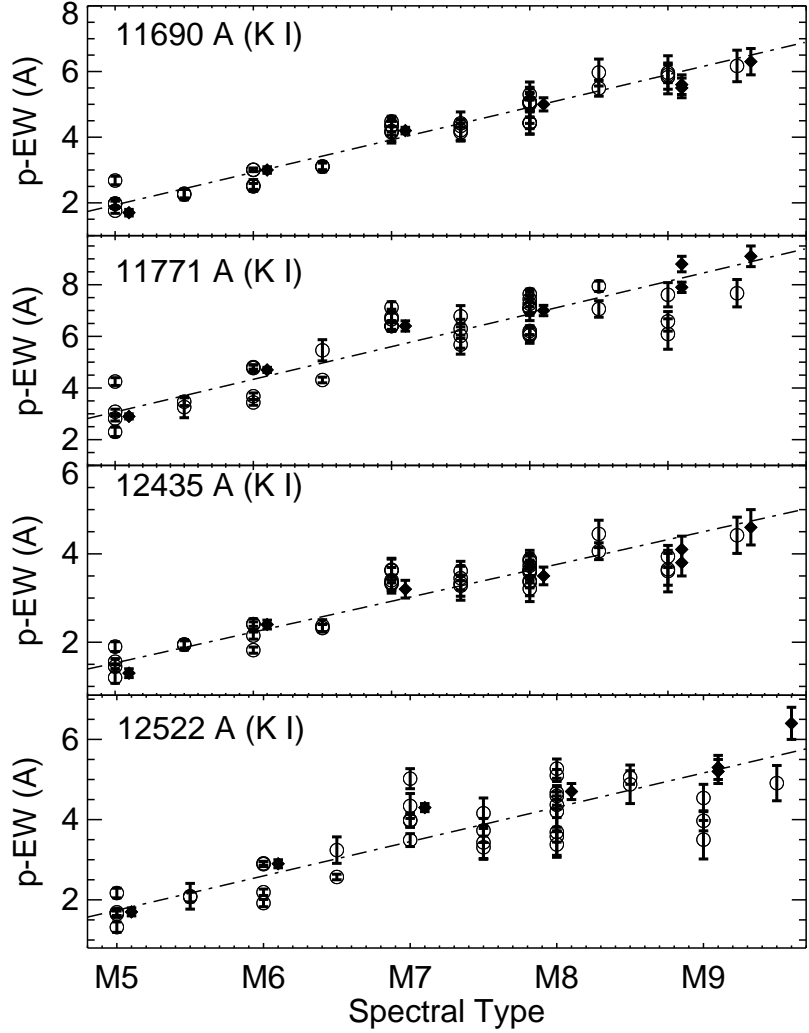}
\end{center}
\caption{Plot of K~I lines in the spectra of M dwarfs. The top two panels are the K~I lines in order 65 while the bottom two are from order 61. Here we compare our work (open circle) with that of \citet{cushing05} (filled diamond, moved to the right by 0.1 spectral type for clarity). The line fitting these points is a linear regression whose equation is listed in Table 6. We find that our results are consistent with those in literature.}
\end{figure}

\clearpage
\begin{figure}
\begin{center}
\includegraphics[width=12.74 cm]{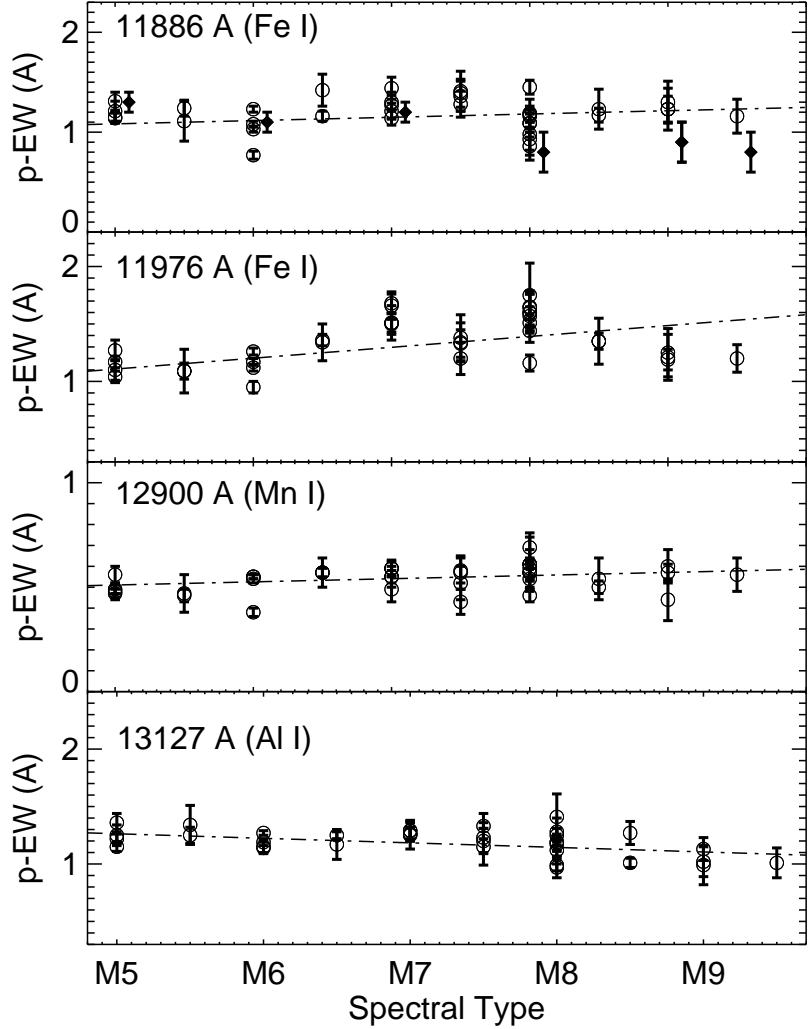}
\end{center}
\caption{The top two panels in this figure are the Fe~I  observed in order 64. The top panel (Fe I at 11890 $\AA$) also compares our work (open circle) with that of \citet{cushing05} (filled diamond, moved to the right by 0.1 spectral type for clarity). Our results agree well within the errors. The bottom plot shows a weak Al~I llnes from order 58. We note that Fe~I (11886 $\AA$) and Mn~I (12900 $\AA$) show little variation with spectral type. The line fitting these points is a linear regression whose equation is listed in Table 7.}
\end{figure}
\clearpage
\begin{figure}[t]
\begin{center}
\includegraphics[width=12.74 cm]{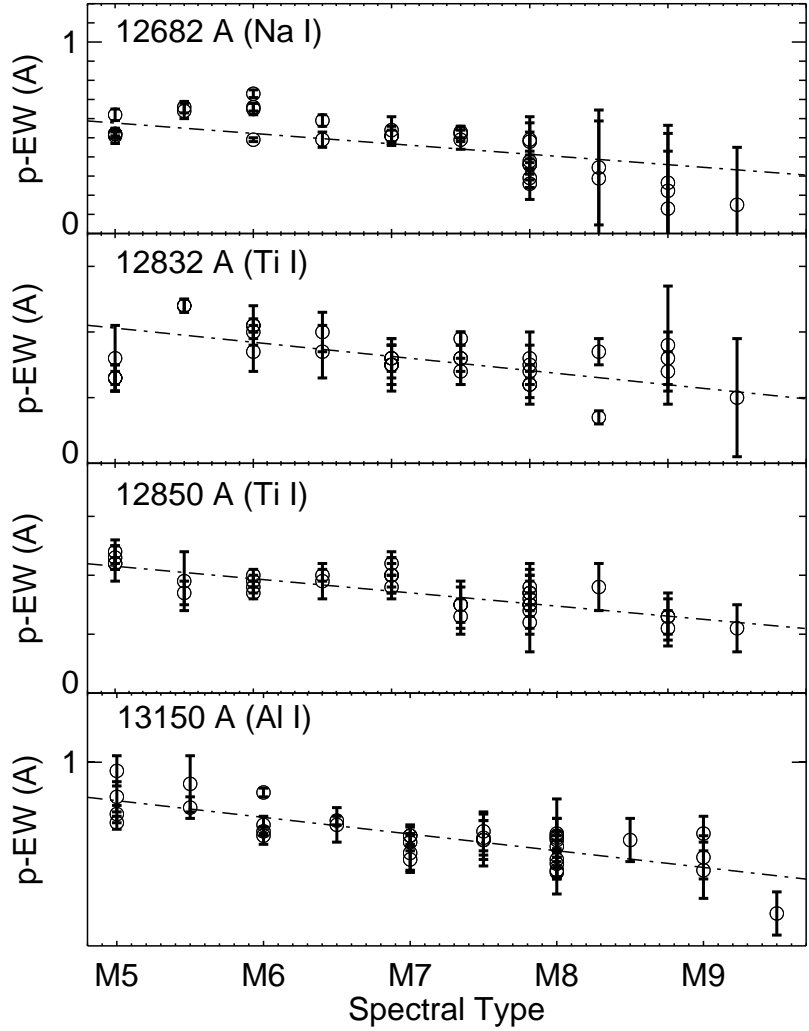} 
\end{center}
\caption{This figure shows a collection of weak lines in M dwarf spectra. The legend in this figure is the same as in figure 6. The line fitting these points is a linear regression whose equation is listed in Table 7. Na I after M8.0 is not observed in the spectra as it is weak and therefore drowned by the noise. Therefore, the p-EW measurements plotted past M8.0 for Na I have large uncertainties.}
\end{figure}

\bibliographystyle{amsplain}
\bibliography{myreferences}
\end{document}